\documentclass[aps,pra,twocolumn,nopacs,floatfix]{revtex4-2}
\usepackage{epsfig}
\usepackage{graphicx}
\usepackage{dcolumn}
\usepackage{amsthm,amsmath}
\usepackage{comment}
\usepackage[colorlinks=true, allcolors=blue]{hyperref}
\usepackage{xcolor}
\usepackage{mathtools}
\usepackage{orcidlink}
\usepackage{amsmath}
\usepackage[normalem]{ulem}
\usepackage{bm}
\usepackage[T1]{fontenc}

\begin{document}

\preprint{APS/1-Knew}

\title{Testing for isospin symmetry breaking with extensive calculations of isotope shift factors in potassium}

\author{$^{a,b}$Vaibhav Katyal}
\author{$^{a,b}$A. Chakraborty~\orcidlink{0000-0001-6255-4584}}

\author{$^a$B. K. Sahoo~\orcidlink{0000-0003-4397-7965}}
\email{bijaya@prl.res.in}

\affiliation{
$^a$Atomic, Molecular and Optical Physics Division, Physical Research Laboratory, Navrangpura, Ahmedabad 380009, India}  
\affiliation{
$^b$Indian Institute of Technology Gandhinagar, Palaj, Gandhinagar 382355, India
}

\author{Ben Ohayon~\orcidlink{0000-0003-0045-5534}}
\email{bohayon@technion.ac.il}
\affiliation{
The Helen Diller Quantum Center, Department of Physics,
Technion-Israel Institute of Technology, Haifa, 3200003, Israel
}

\author{Chien-Yeah Seng~\orcidlink{0000-0002-3062-0118}}
\affiliation{Facility for Rare Isotope Beams, Michigan State University, East Lansing, MI 48824, USA}
\affiliation{Department of Physics, University of Washington,
	Seattle, WA 98195-1560, USA}

\author{Mikhail Gorchtein}
\affiliation{Institut f\"ur Kernphysik, Johannes Gutenberg-Universit\"at Mainz, 55128 Mainz, Germany}
\affiliation{
PRISMA$^+$ Cluster of Excellence, Johannes Gutenberg-Universit\"at Mainz, 55128 Mainz, Germany
}

\author{John Behr}
\affiliation{TRIUMF, 4004 Wesbrook Mall, Vancouver, British Columbia, Canada V6T 2A3}

\date{\today}

\begin{abstract}
Precise evaluation of the isotope shift (IS) factors for seven low-lying potassium (K) states is achieved using relativistic coupled-cluster (RCC) theory.
The energies of these states are assessed and compared with experimental data to confirm the accuracy of the wave functions calculated at varying RCC theory approximations and highlight the significance of many-body and relativistic effects in determining the energies and IS factors of K.
Various methods are used to compute the IS factors, with the finite-field (FF) approach yielding results that align with observed and semi-empirical data. This consistency is attributed to orbital relaxation effects that are naturally present in the FF method but emerge only through complex interactions in other techniques.
Using the IS factors derived from FF, we review the mean square radius difference between $^{38m}$K and $^{39}$K. From this difference and muonic atom x-ray spectroscopy, we deduce the absolute radius of $^{38m}$K using an updated calculation of the nuclear polarizability effect. Finally, we evaluate the isospin symmetry breaking (ISB) in this isotriplet by integrating the radius of $^{38m}$K with an updated radius of $^{38}$Ca, concluding that the ISB is compatible with zero.
This finding offers a stringent benchmark for nuclear model calculations of ISB corrections in nuclear beta decay, which play a key role in determining the $V_{ud}$ matrix element.
\end{abstract}
            
\maketitle

\section{\label{sec1} Introduction}

The change in nuclear charge radii among the isotopes of an element can be inferred precisely from isotope shift (IS) studies~\cite{2011-Blaum,2013-Blaum,King,2023-exprev}.
These changes may then be combined with the absolute radius of some reference isotope, extracted from muonic atom x-ray spectroscopy~\cite{1969-Muonic,1974-Engfer,2020-Nuc,2022-Muonic}, to obtain the absolute radii of the entire chain.

Knowledge of the absolute charge radii of the spin $J=0$ and isospin $T=1$ isobaric states is important for the precise determination of the Cabibbo-Kobayashi-Maskawa (CKM) \cite{Cabibbo:1963yz,Kobayashi:1973fv} matrix element $V_{ud}$ from superallowed nuclear beta decays~\cite{Hardy:2020qwl}.
Recently, an observed $\sim 3\sigma$ deficit in the CKM unitarity relation of the top row $|V_{ud}|^2+|V_{us}|^2+|V_{ub}|^2=1$~\cite{Cirigliano:2022yyo} has triggered renewed interest in the nuclear and particle physics communities and is currently under careful scrutiny~\cite{Gorchtein:2023naa}.
An improved extraction of $V_{ud}$ from the superallowed decays requires a more accurate input from the SM; among them is the ``statistical rate function'' $f$~\cite{Seng:2022inj,Seng:2023cgl} and the isospin symmetry breaking (ISB) correction $\delta_\text{C}$~\cite{Seng:2022epj,Seng:2023cvt,Plattner} which benefit the most from the more accurate determination of the nuclear charge radii across the superallowed isotriplet.

The $A=38$ nuclear isotriplet ($^{38}$Ca, $^{38m}$K, $^{38}$Ar) is the only one in which all three charge radii are experimentally measured. On the one hand, with the help of isospin symmetry, the charge radii allow pinning down the so-called shape factor which describes the spatial distribution of the decay probability inside the decaying nucleus~\cite{Seng:2022inj,Seng:2023cgl}. On the other hand, exact isospin symmetry implies a certain pattern that the nuclear radii in the isotriplet must follow. The measurement of deviations from this pattern provides a direct test of ISB effects~\cite{Seng:2022epj}. 

Extracting the charge radii of unstable nuclei such as $^{38m}$K from high-precision isotope-shift measurements typically requires high-precision calculations of IS factors~\cite{review}.
This is particularly pronounced in elements with an odd number of protons, such as K, where only up to two naturally abundant isotopes are available~\cite{2012-Coc}.
Many of the low-lying states of this atom have a closed core $[3p^6]$ and a valence electron in the $s$ and $p$ orbitals. Therefore, the electron correlation effects in this system are easy to understand, and it is possible to calculate the IS factors of these low-lying states of K accurately using modern state-of-the-art methods on a high-performance cluster (HPC).  In addition, muonic $^{39,41}$K measurements are also available to perform a comparative analysis between electronic and muonic data~\cite{Ehrlich}.

To our knowledge, at least four different groups have reported {\it ab initio} calculations of IS factors for a few low-lying states of the K atom using various many-body methods~\cite{Martensson1,Martensson2,Safronova,Berengut1,Dzuba1,Sourav}.
However,the results obtained from all these calculations show large deviations from each other. In the 1990s, Martensson and collaborators have performed a series of calculations on the field-shift (FS) factors and hyperfine structure factors in the ground state and $4P$ state using the coupled-cluster (CC) method in the nonrelativistic approximation. They had included approximate relativistic corrections to improve the accuracy of the results, which were found to be significant~\cite{Martensson1,Martensson2}.
In these calculations, an expectation value evaluation (EVE) approach was employed to account for the electron correlation effects on the determination of these factors via Goldstone diagram representation. In 2001, Safronova and Johnson extended this approach to evaluate the FS and specific mass shift (SMS) factors of seven low-lying states of K using a mixture of singles and doubles approximated linearized CC theory (SD method), random phase approximation (RPA) and many-body perturbation theory (MBPT) in the relativistic framework~\cite{Safronova}. Following this work, Berengut \textit{et al.} employed a finite field (FF) approach through a correlation operator that accounts for the Brueckner effects in the Dirac-Hartree-Fock (DHF) method to estimate the FS and SMS factors of the $4P_{1/2}-4S$ and $4D_{3/2}-4S$ transitions of K~\cite{Berengut1}. They had also multiplied a scaling factor with the correlation operator in order to fit the calculated energies with the experimental values. A decade later, the calculations were revised using the FF approach employing the relativistic CC (RCC) method~\cite{Sourav}. These calculations were significantly different from the previous ones. 

The preceding computations employed the non-relativistic formulation for the SMS operator. The most recent evaluations of the IS factors involved estimating the FS and SMS factors for the $D_1$ line of K using the analytical response (AR) method within the framework of the RCC theory~\cite{Koszorus}. Here, the relativistic formulation of the SMS operator was employed, but the specifics of these calculations were not provided.

In this work, we present first-principle calculations of the FS, normal mass shift (NMS), and SMS factors of the $4S$, $4P_{1/2;3/2}$, $5S$, $5P_{1/2;3/2}$ and $6S$ states of K.
The calculations employ FF, EVE and AR methods with RCC theory to examine how the chosen approach impacts the results.
We have also analyzed the trends of the results in the DHF, MBPT(2), singles and doubles approximated RCC theory (RCCSD method), and singles, doubles and triples approximated RCC theory (RCCSDT method) from all the three approaches. Furthermore, we have calculated the energies and magnetic dipole hyperfine structure factors ($A_{hf}$) of the above states of K using RCC theory. Comparison of these values with the available experimental values can help to understand the precision of the calculated wave functions. To demonstrate the importance of relativistic corrections, we have calculated the energies and IS factors by considering a large value for the speed of light ($c$).

We compare the calculations with each other and perform various tests against electronic and muonic measurements to arrive at a robust estimation of the uncertainty of IS factors.
Using the chosen factors and new macroscopic calculations of the nuclear polarizability, we update the radii of $^{38m}$K and $^{38}$Ca. These serve as a crucial input for an accurate evaluation of the $ft$ values of the superallowed decays in the $A=38$ isotriplet \cite{Seng:2023cgl}, where $t$ is the half-life. From these radii and the results of the mirror fit of Ref.~\cite{2024-Mirr} we construct the ISB test quantity $\Delta M_B^{(1)}$ which benchmarks the ISB calculations needed for extracting $V_{ud}$ from the $ft$ values of superallowed decays.

\section{Theory}

For an atomic energy level $i$, the first-order IS $\delta E_i^{A,A'}$ between elements $A$ and $A'$ can be approximated as~\cite{bohr1922difference, 1922-Ehr}
\begin{equation}
\delta E_i^{A,A'} = F_i~\delta \langle r_N^2 \rangle^{A,A'} + K_i^{\text{MS}}(\mu_{A'}-\mu_A), 
\label{IS}
\end{equation}
where $F$ and $K^{\text{MS}}$ are known as the FS and mass shift (MS) factors, respectively, which can be determined by performing atomic calculations of their respective operators (see, e.g.,~\cite{review}). $\delta \langle r_N^2 \rangle^{A,A'}\equiv\langle r_N^2 \rangle^{A'}-\langle r_N^2 \rangle^{A}$ is the change in the mean square nuclear radius between the two isotopes, and $\mu_A=(M_A+m_e)^{-1}$ with $m_e$ the electron mass and $M_A$ the nuclear mass.

The MS factor can be parametrized as $K^{\text{MS}}=K^{\text{NMS}} +K^{\text{SMS}}$. $K^{\text{NMS}}$ corresponds to the contribution of the NMS operator, which is the one-body part of the total recoil operator. 
$K^{\text{SMS}}$ corresponds to the contribution of the SMS operator, which is the two-body part of the total recoil operator. 

The FS operator is given as~\cite{review}
\begin{eqnarray}\label{eq:Fop}
\bar{F} = \sum_i^{N_e} f (r_i) = - \sum_i^{N_e} \frac{\delta V_n(r_i)^{A',A}}{\delta \langle r_N^2 \rangle^{A',A} } ,
\end{eqnarray}
where $N_e$ is the number of electrons, and $\delta V_n(r_i)^{A,A'}$ is the change in the nuclear potential ($V_n(r)$) between isotopes $A$ and $A'$.
The FS factor $F$ can be calculated from its respective operator $\bar{F}$ using atomic theory methods. 
Alternatively, for $ns$ states, $F$ may be calculated semi-empirically by relating it to the magnetic dipole hyperfine structure factor via~\cite{Martensson2}
\begin{eqnarray}
F = - D Z A_{hf}/ g_I ,
\label{eqFA}
\end{eqnarray}
where $Z$ is the atomic number, $g_I= \mu_I/I$ is the nuclear g-factor, $\mu_I$ is the magnetic moment, $I$ is the nuclear spin, and $D\approx6.16381 \times 10^{-3}~$fm$^{-2}$. 

The precise calculations of $F$ and $A_{hf}$  depend on the accurate determination of the atomic wave functions in the nuclear region. Thus, it is useful to calculate the $A_{hf}$ values and compare them with their measured values to assess the precision of the $F$ factors.
$A_{hf}$ is given by~\cite{charles}
\begin{eqnarray}
A_{hf} = \mu_N g_I \frac{\langle J || T_{hf}^{(1)}|| J\rangle}{\sqrt{J(J+1)(2J+1)}} ,
\end{eqnarray}
where $\mu_N$ is the nuclear magneton, $J$ is the angular momentum of the state and $T_{hf}^{(1)}$ is magnetic dipole hyperfine structure operator
\begin{eqnarray}\label{eq:Aop}
T_{hf}^{(1)} = \sum_k t_q^{(1)}(r_k) \equiv \sum_k - \iota e \sqrt{\frac{8 \pi}{3}} \frac{{\vec \alpha} \cdot {\vec Y}^1_q({\hat r}_i)}{r_i^2}.
\end{eqnarray}

The relativistic forms of the NMS and SMS operators, up to the order of $\alpha Z$, in atomic units (a.u.) are given by~\cite{Shabaev}
\begin{eqnarray}\label{nmsexp}
O^{\text{NMS}} &=& \sum_i^{N_e} o_i^\text{NMS} \nonumber \\
&\equiv&\frac{1}{2}\sum_i^{N_e} \left ({\vec p}_i^{~2} - \frac{\alpha Z}{r_i} {\vec \alpha}_i^D \cdot {\vec p}_i 
- \frac{\alpha Z}{r_i} ({\vec \alpha}_i^D \cdot {\vec C}_i^1){\vec C}_i^1 \cdot {\vec p}_i \right ) \nonumber 
\end{eqnarray}
and 
\begin{eqnarray}\label{smsexp}
\hspace{-13mm}
O^{\text{SMS}} = \sum_{i \ne j}^{N_e} o_{ij}^\text{SMS} 
&\equiv& \frac{1}{2} \sum_{i\ne j}^{N_e} \left ({\vec p}_i \cdot {\vec p}_j - \frac{\alpha Z}{r_i} {\vec \alpha}_i^D \cdot {\vec p}_j 
 \right. \nonumber \\ && \left. - \frac{\alpha Z}{r_i} ({\vec \alpha}_i^D \cdot {\vec C}_i^1) ({\vec p}_j \cdot {\vec C}_j^1) \right ), 
\end{eqnarray}
respectively, where ${\vec p}$ is the momentum operator, ${\vec \alpha}_i^D $ is the Dirac matrix, ${\vec C}^1$ is the Racah operator, and $\alpha$ is the fine-structure factor. The last two terms in the above expressions are the leading relativistic corrections. When these terms are dropped, the expressions correspond to their non-relativistic forms. 

In the relativistic framework, the single-particle electron wave function $|\phi \rangle$ is given by
\begin{eqnarray}
|\phi \rangle &=& \frac{1}{r} \begin{pmatrix} 
     P(r) & \chi_{\kappa, m_j}(\theta, \phi) \\
     \iota Q(r) & \chi_{-\kappa, m_j}(\theta, \phi) 
   \end{pmatrix}  ,
\end{eqnarray}
where $P(r)$ and $Q(r)$ are the large and small components of the radial part of the wave function, respectively, and $\chi_{\kappa, m_j}(\theta, \phi)$ is the angular factor with the relativistic angular momentum quantum number $\kappa$ and azimuthal component $m_j$ of the total angular momentum $j$. Using these wave functions, the single-particle matrix elements of the different operators can be estimated.
The FS operator of Eq.~(\ref{eq:Fop}) results in
\begin{eqnarray}
\langle \phi_f | f | \phi_i \rangle 
&=&  \delta(\kappa_f,\kappa_i) \delta(m_{jf}, m_{ji}) \times \nonumber \\
&&\int_0^{\infty} dr  f(r) \left ( P_f(r) P_i (r) + Q_f(r) Q_i (r) \right ), \ \  
\end{eqnarray} 
the magnetic dipole hyperfine structure operator of Eq.~(\ref{eq:Aop}) returns
\begin{eqnarray}
\langle \phi_f | t_q^{(1)} | \phi_i \rangle &=& (-1)^{j_f+m_f} \begin{pmatrix} 
     j_f & 1 & j_i \\
     -m_f & q & m_i 
   \end{pmatrix}  (-1)^{j_f+1/2} \nonumber \\ 
  \times &&  \sqrt{(2j_f+1)(2j_i+1)} \begin{pmatrix} 
     j_f & 1 & j_i \\
     \frac{1}{2} & 0 & -\frac{1}{2} 
   \end{pmatrix} (\kappa_f + \kappa_i) \nonumber \\
   \times &&  \int_0^{\infty} dr \frac{P_f(r)Q_i(r)+Q_f(r)P_i(r)}{r^2}, 
\end{eqnarray} 
and the NMS operator of Eq.~(\ref{nmsexp}) returns~\cite{Joensson}
\begin{eqnarray}
\langle \phi_f | o^\text{NMS} | \phi_i \rangle &=& \frac{1}{2} \delta(\kappa_f,\kappa_i) \delta(m_{jf}, m_{ji}) \nonumber \\ & \times & \int_0^{\infty} dr  \left ( \frac{\partial P_f(r)}{\partial r} \frac{\partial P_i(r)}{\partial r}  \right. \nonumber \\ 
&+& \left. \frac{\partial Q_f(r)}{\partial r} \frac{\partial Q_i(r)}{\partial r} +  \frac{l_i (l_i+1) P_f(r) P_i(r)} {r^2} \right. \nonumber \\ 
&+& \left. \frac{\tilde{l}_i (\tilde{l}_i+1) Q_f(r) Q_i(r) }{r^2} - 2 \frac{\alpha Z}{r} \right. \nonumber \\ &\times & \left. \left ( Q_f(r)\frac{\partial P_i(r)}{\partial r} + \frac{\partial P_f(r)}{\partial r} Q_i(r) \right ) -  \frac{\alpha Z}{r^2}  \right. \nonumber \\ &\times&  \left. (\kappa_i -1) \left ( Q_f(r) P_i(r) + P_f(r) Q_i(r) \right )  \right ) , \ \ \ \ \
\label{relnms}
\end{eqnarray} 
where $l$ and $\tilde{l}$ are the orbital radial quantum numbers for the large and small components, respectively. 

The SMS operator of Eq.~(\ref{smsexp}) is a scalar product of two rank-one operators. Thus, its single particle matrix element can be expressed as~\cite{Joensson}
\begin{eqnarray}
\langle \phi_i \phi_j | o^\text{SMS} | \phi_k \phi_l \rangle &=& \delta(m_{ji}-m_{jk}, m_{jl}-m_{jj})  \nonumber \\ 
&& \times \sum_{q=-1}^{1} \begin{pmatrix} 
     j_i & 1 & j_k \\
     -m_i & q & m_k 
   \end{pmatrix} \begin{pmatrix} 
     j_j & 1 & j_l \\
     -m_j & -q & m_l 
   \end{pmatrix} \nonumber \\ 
&& \times (-1)^{j_i-m_i + j_j -m_j +1 -q} X^1(ij,kl) ,
\end{eqnarray}
where $X^1(ij,kl)$ is the reduced matrix element given by
\begin{eqnarray}
X^1(ij,kl) &=& \sqrt{(2j_i+1) (2j_j+1) (2j_k+1) (2j_l+1)} \nonumber \\
&& \times \begin{pmatrix} 
     j_i & 1 & j_k \\
     1/2 & 0 & -1/2 
   \end{pmatrix} \begin{pmatrix} 
     j_j & 1 & j_l \\
     1/2 & 0 & -1/2 
   \end{pmatrix} \nonumber \\
   && \times (-1)^{j_i+j_j+1} \left [ R(ik) R(jl) + \frac{1}{2} \left ( R(ik)X(jl) \right. \right. \nonumber \\ 
   && \left. \left. +X(ik)R(jl) \right ) \right ]
\label{relsms}
\end{eqnarray}
with the radial functions
\begin{eqnarray}
 R(ab) &=& - \iota \int_0^{\infty} dr \left [ P_a(r) \left ( \frac{\partial P_b(r)}{\partial r} \right. \right. \nonumber \\ &-& \left. \left. \frac{\kappa_a(\kappa_a-1)-\kappa_b(\kappa_b-1)}{2r} P_b(r) \right ) + Q_a(r) \left ( \frac{\partial Q_b(r)}{\partial r} \right. \right. \nonumber \\ &-& \left. \left.   \frac{\kappa_a(\kappa_a+1)-\kappa_b(\kappa_b+1)}{2r} Q_b(r) \right ) \right ] 
\end{eqnarray}
and
\begin{eqnarray}
 X(ab) &=& - \iota \int_0^{\infty} dr \frac{\alpha Z}{r} \left [ (\kappa_a - \kappa_b -2) P_a(r) Q_b(r) \right. \nonumber \\
 && \left. + (\kappa_a - \kappa_b  + 2) Q_a(r) P_b(r)\right ] . 
\end{eqnarray}

\section{Method of calculation}

\subsection{General description}

In the infinite nuclear mass limit, the general form of the atomic Hamiltonian reads
\begin{eqnarray}
H_\text{at} = \sum_i h(r_i) + \sum_{i,j>i} g(r_{ij}) ,
\end{eqnarray}
where $h$ and $g$ denote, respectively, the one-body and two-body interactions at the single particle level. Due to the presence of two-body interactions, it is not possible to solve the wave function of the above Hamiltonian directly.
This problem is addressed by using a mean-field approach (MF), in which the Hamiltonian is approximated as an effective one-body operator $H_0$ to obtain an approximated wave function ($|\Phi_0 \rangle$) of an atomic state. A residual interaction, $V_\text{res}=H_\text{at}-H_0$, is neglected at this stage. It follows that the effective Hamiltonian contains all the one-body terms of $H_{at}$ and an effective one-body potential ($U_\text{MF}=\sum_i u_\text{MF}(r_i)$) constructed from $g$. i.e. $H_0= \sum_i h_0(r_i) = \sum_i (h(r_i) + u_\text{MF}(r_i))$.

By applying the variational principle, the MF potential for single particle wave function $|\phi_i \rangle$ of $|\Phi_0 \rangle$ reads
\begin{eqnarray}
u_\text{MF} |\phi_i \rangle&=& \sum_{a=1}^{N_c} \left [ \langle \phi_a | g | \phi_a \rangle |\phi_i \rangle - \langle \phi_a | g | \phi_i \rangle | \phi_a \rangle \right ] ,
\end{eqnarray}
where $N_c$ is the number of occupied orbitals in the system and $|\phi_c \rangle$s are their wave functions.
Thus, the single-particle orbital-determining equation, $h_0 |\phi_i \rangle = \epsilon_i |\phi_i \rangle$ with orbital energy $\epsilon_i$, is written as
\begin{eqnarray}
h |\phi_i \rangle + \sum_{a=1}^{N_c} \left [ \langle  \phi_a | g | \phi_a \rangle | \phi_i \rangle - \langle \phi_a | g | \phi_i \rangle | \phi_a \rangle \right ] = \epsilon_i |\phi_i \rangle . \ \ \ \
\end{eqnarray}
It follows that the single-particle orbital energy expression is
\begin{eqnarray}
\epsilon_i = \langle \phi_i | h |\phi_i \rangle + \sum_{a=1}^{N_c} \left [ \langle \phi_i \phi_a | g | \phi_i \phi_a \rangle - \langle \phi_i \phi_a | g | \phi_a \phi_i \rangle \right ] . \ \ \ \ \
\label{eqdfeng}
\end{eqnarray}

In this work, we begin with $H_\text{at}$ as the Dirac-Coulomb (DC) Hamiltonian to calculate atomic wave functions and energies in the relativistic framework, given in a.u., by
\begin{eqnarray}\label{eq:DC}
H_\text{at} &=& \sum_i \left [c {\vec \alpha}_i^D \cdot {\vec p}_i+(\beta_i^D-1)c^2+V_n(r_i)\right] \nonumber \\ && +\sum_{i,j>i}\frac{1}{r_{ij}}, 
\end{eqnarray}
where ${\vec \alpha}^D$ and $\beta^D$ are the Dirac matrices, ${\vec p}$ is the single particle momentum operator and $\frac{1}{r_{ij}}$ represents the Coulomb potential between the electrons. We assume the Fermi charge distribution in our calculations unless otherwise stated explicitly. Corrections due to the Breit interactions and lower-order QED effects~\cite{bijaya1} are subsequently added to the DC contributions.   
The Breit interaction is accounted on par with the two-body Coulomb interaction term in a self-consistent manner at the DHF and RCC methods. The lowest order vacuum polarization and self-energy QED effects are also incorporated self-consistently in the generation of DHF orbitals, but these interactions are defined using model potentials~\cite{bijaya1,ginges}. Thus, the estimated QED corrections in this work are not reliable enough and represent only typical order-of-magnitudes neglected because of the omission of QED interactions. We account for them as a part of the total uncertainty to the final result along with contributions from other error sources as discussed later. 

In this work, we calculate the energies and IS factors of the $4S$, $4P_{1/2;3/2}$, $5S$, $5P_{1/2;3/2}$ and $6S$ states of K. All these states have a common closed-shell configuration $[3p^6]$ and differ by a valence orbital (denoted by $v$). To determine the wave functions, we first calculate the MF wave function, $|\Phi_0 \rangle$, of the common closed-core reference $[3p^6]$ using the DHF method.
Then, the exact wave function of the closed-core is obtained by including the correlation effects due to $V_\text{res}$. In the next step, we determine the wave function of a state of K with exact configuration by appending the corresponding valence orbital, $v$, of the respective state in the Fock-space approach.

With knowledge of the DHF wave function $|\Phi_0 \rangle$, the exact wave function $|\Psi_0 \rangle$ of the closed core $[3p^6]$ can be determined by operating a wave operator $\Omega_0$ on $|\Phi_0 \rangle$. i.e.
\begin{eqnarray}
|\Psi_0 \rangle = \Omega_0 |\Phi_0 \rangle .
\end{eqnarray}
The generalized Bloch equation which prescribes the procedure to determine the amplitude of $\Omega_0$ and the energy of the corresponding state, is given by~\cite{Lindgren1985}
\begin{eqnarray}
[\Omega_0,H_0] |\Phi_0 \rangle = [V_\text{res} -E_{corr}] \Omega_0 |\Phi_0 \rangle .
\label{eqbloch}
\end{eqnarray}
Here, $E_{corr}=\langle \Phi_0 | V_\text{res} \Omega_0 |\Phi_0 \rangle$ is the correlation energy that contributes to the total energy of the ground state, $E_0= E_{DF} + E_{corr}$ with $E_{DF}=\langle \Phi_0 | H_0 |\Phi_0 \rangle$. 

The DHF wave function of the actual state of interest of K with a valence orbital $v$ can be defined as $|\Phi_v \rangle = a_v^{\dagger} |\Phi_0 \rangle$ in the V$^{N_c-1}$ potential approximation. Then, the exact wave function, $|\Psi_v \rangle$, can be obtained by~\cite{Lindgren1985,bijaya2,bijaya3}
\begin{eqnarray}
|\Psi_v \rangle = (\Omega_0 + \Omega_v) |\Phi_v \rangle ,
\end{eqnarray}
where $\Omega_0$ is responsible for accounting for electron correlation effects due to core orbitals (it is practically the same as the one defined for $|\Psi_0 \rangle$) and $\Omega_v$ takes care of correlation effects of the valence electron with the core electrons, respectively, due to the residual interaction $V_\text{res}$. Thus, it follows that after solving the amplitudes of $\Omega_0$, the amplitudes of $\Omega_v$ can be determined by substituting $\Omega_0+\Omega_v$ into the generalized Bloch equation given by Eq.~(\ref{eqbloch}).
It yields
\begin{eqnarray}
 [\Omega_v, H_0] |\Phi_v \rangle &=& V_\text{res} (\Omega_0 + \Omega_v) |\Phi_v \rangle  -  E_v \Omega_v |\Phi_v \rangle ,
\label{blwv}
\end{eqnarray}
where $E_v$ is the net energy of the state and evaluated as  
\begin{eqnarray}
E_v = \langle\Phi_v| V_\text{res} (\Omega_0 + \Omega_v) |\Phi_v\rangle.
\end{eqnarray}
Here, a normal-order Hamiltonian with respect to the reference state $|\Phi_0 \rangle$ is used in the calculation. In such a case, $E_v$ corresponds to the electron attachment energy (AE) and is equivalent to negative of the ionization potential of the state.

\subsection{Evaluation of IS factors}

\subsubsection{Different approaches}

We calculate IS factors related to the first-order shifts to energy levels of $H_\text{at}$ due to the FS, NMS and SMS interactions.
The first order energy shift of an atomic state can be evaluated in three different approaches: (i) in the FF approach, in which the total energy of the state is expanded to estimate the first order energy as the first-order derivative, (ii) by evaluating the expectation value of the interaction Hamiltonian in the EVE approach and (iii) by solving the Schr\"odinger equation perturbed in first order~\cite{bijaya4,review}.

In the FF approach, we define the new Hamiltonian as $H=H_\text{at}+\lambda_o O$ to estimate the IS factors with $O=\sum_i o(r_i)$ denoting the FS or MS operators with $\lambda_o$ as an arbitrary parameter with a square of radii dimension or inverse mass, respectively. For brevity, the calculated energies and wave functions of $H$ are presented without any superscript, while the energies and wave functions of $H_\text{at}$ are denoted by superscript ``$(0)$'' and their first-order corrections due to the interaction Hamiltonian $O$ are denoted by superscript ``$(1)$''.        

The $\lambda_o$-dependent energy, $E_v(\lambda_o)$, of an atomic state $|\Psi_v \rangle$ due to $H$ can be calculated in the FF approach by allowing a small value of $\lambda_o$ so that it can be expressed as 
\begin{eqnarray}
E_v(\lambda_o) &=& E_v^{(0)} + E_v^{(1)} + E_v^{(2)} + \cdots \nonumber \\
    &=& E_v^{(0)} + \lambda_o \langle O \rangle + {\cal{O}} \left ( \lambda_o^2\right) ,
\end{eqnarray}
where $\langle O \rangle$ corresponds to the respective IS factor due to $O$. The value of $\lambda_o$ is chosen to be small enough so that terms with higher powers of $\lambda_o$ can be safely neglected in the above expansion.
The IS factor can then be extracted from the calculated $E_v(\lambda_o)$ values as
\begin{eqnarray}
\langle O \rangle \simeq \frac{E_v(+\lambda_o) - E_v(-\lambda_o)}{2 \lambda_o} .
\label{eqff}
\end{eqnarray}

In principle, the optimal choice of $\lambda_o$ can be state dependent as the magnitude of $O$ can vary by one or two orders for different states. However, for practical reasons, we have considered the same value $\lambda_o=10^{-5}$~a.u. for all states. 

The aforementioned issues of $\lambda_o$ dependencies in the estimation of the IS factors in the FF approach can be removed by evaluating the expectation value of $O$ directly in the EVE approach as
\begin{eqnarray}
 \langle O \rangle = \frac{\langle \Psi_v^{(0)} | O | \Psi_v^{(0)} \rangle}{\langle \Psi_v^{(0)} | \Psi_v^{(0)} \rangle} .
 \label{eqeve}
\end{eqnarray}
It requires the determination of a normalization factor for the wave function. Moreover, in the RCC theory framework, both numerator and denominator of the EVE expression contain non-terminating series, as shown later.

The AR approach circumvents the issues of both the FF approach and the EVE approach. To derive the IS expression of the AR approach, we expand the wave function of the atomic state as
\begin{eqnarray}
|\Psi_v \rangle &=&  |\Psi_v^{(0)} \rangle + \lambda_o |\Psi_v^{(1)} \rangle  + \cdots .
\end{eqnarray}
The first-order wave-function-solving equation is given by
\begin{eqnarray} 
 \left ( H_\text{at} - E_v^{(0)}\right ) |\Psi_v^{(1)} \rangle = \left ( \langle O \rangle - O \right ) |\Psi_v^{(0)} \rangle .
 \label{eqar}
\end{eqnarray}
It is evident from the above discussions that the AR approach is the more favorable approach to estimating the IS factors. However, we learn from the present study that this may not always be the case, particularly in K, because of the approximations made in the determination of atomic wave functions. 

\subsubsection{Employed methods}

The methods described above for calculating the IS factors can be implemented with varying degrees of approximation in many-body methods.
Developing a many-body approach begins with establishing a mean-field wave function, which is formed through the DHF method. In this study, IS factors derived from DHF wave functions are referred to as DHF values.

To estimate the IS factors in the EVE [Eq.~(\ref{eqeve})] and AR [Eq.~(\ref{eqar})] approaches, unperturbed wave functions are used. This gives the same DHF value for an IS factor in both approaches.
The following expression for the FS or NMS factors, due to the respective one-body operators $O^\text{FS/NMS}$, using the unperturbed DHF wave function $|\Phi_v^{(0)} \rangle$ is
\begin{eqnarray} \label{dhffn}
\langle O^\text{FS/NMS} \rangle &=& \frac{\langle \Phi_v^{(0)} | O^\text{FS/NMS} | \Phi_v^{(0)} \rangle}{\langle \Phi_v^{(0)} | \Phi_v^{(0)} \rangle} \nonumber \\
    &=&  \sum_c^{N_c} \langle \phi_c^{(0)} | o^\text{FS/NMS} | \phi_c^{(0)} \rangle \nonumber \\ && + \langle \phi_v^{(0)} | o^\text{FS/NMS} | \phi_v^{(0)} \rangle .
\end{eqnarray}
The contributions from the first term correspond to the closed $[3p^6]$ core in K are shared among all the states considered in this study and do not impact the differential IS factors of a transition between these states. Therefore, they are omitted when reporting the DHF values of the FS and NMS factors.

Since the SMS operator is a two-body operator, its DHF expression is given by
\begin{eqnarray}\label{dhfs}
\langle O^\text{SMS} \rangle &=& \frac{1}{2} \sum_{ab}^{N_c} \left [ \langle \phi_a^{(0)} \phi_b^{(0)}| o^\text{SMS} | \phi_a^{(0)} \phi_b^{(0)} \rangle \right. \nonumber \\ && \left. - \langle \phi_b^{(0)} \phi_a^{(0)}| o^\text{SMS} | \phi_a^{(0)} \phi_b^{(0)} \rangle \right ] \nonumber \\
&& +  \sum_a^{N_c} \left [ \langle \phi_v^{(0)} \phi_a^{(0)}| o^\text{SMS} | \phi_v^{(0)} \phi_a^{(0)} \rangle \right. \nonumber \\ && \left. - \langle \phi_a^{(0)} \phi_v^{(0)}| o^\text{SMS} | \phi_v^{(0)} \phi_a^{(0)} \rangle \right ]  .
\end{eqnarray}
Similar to the FS and NMS factors, the first two terms correspond to contributions from the closed-core and are not considered. Thus, contributions only from the last term is referred to as the DHF value. It is defined as
\begin{eqnarray}
\langle O^\text{SMS} \rangle &=& \langle \phi_v^{(0)} | o_1^\text{SMS} | \phi_v^{(0)} \rangle .
\end{eqnarray}

In the FF approach, the single particle DHF orbital energy can be expanded as
\begin{eqnarray}
\epsilon_v &=& \epsilon_v^{(0)} + \epsilon_v^{(1)} + \epsilon_v^{(2)} + \cdots \nonumber \\
           &=&  \epsilon_v^{(0)} + \lambda_o \langle o \rangle + {\cal{O}} \left ( \lambda_o^2\right) ,  
\end{eqnarray}
where $\langle o \rangle= \langle \phi_v^{(0)} | o | \phi_v^{(0)} \rangle$ is the single particle contribution to the IS factor to the valence orbital. Following Eq. (\ref{eqff}), $ \langle o \rangle$ can be estimated by
\begin{eqnarray}
\langle o \rangle \simeq \frac{\epsilon_v(+\lambda_o) - \epsilon_v(-\lambda_o)}{2 \lambda_o} ,
\label{eqffdf}
\end{eqnarray}
and it can be referred to as the DHF value of the IS factor in the FF approach. Eq.~(\ref{eqdfeng}) for the FS and NMS factors yields
\begin{eqnarray}
\langle o^\text{FS/NMS} \rangle &=& \langle \phi_v^{(0)} | o^\text{FS/NMS} |\phi_v^{(0)} \rangle +  \langle \phi_v^{(0)} | h_0 |\phi_v^{(1)} \rangle  \nonumber \\
 +\sum_{a=1}^{N_c}&&\left [ \langle \phi_v^{(0)} \phi_a^{(0)} | g | \phi_v^{(1)} \phi_a^{(0)} \rangle - \langle \phi_v^{(0)} \phi_a^{(0)} | g | \phi_a^{(0)} \phi_v^{(1)} \rangle \right ] \nonumber \\
 +\sum_{a=1}^{N_c}&&\left [ \langle \phi_v^{(0)} \phi_a^{(0)} | g | \phi_v^{(0)} \phi_a^{(1)} \rangle - \langle \phi_i^{(0)} \phi_a^{(0)} | g | \phi_a^{(1)} \phi_v^{(0)} \rangle \right ] . \nonumber \\
\label{eqdfeng1}
\end{eqnarray}
For the SMS factor, it corresponds to
\begin{eqnarray}
\langle o^\text{SMS} \rangle &=& \langle \phi_v^{(0)} | o_1^\text{SMS} |\phi_v^{(0)} \rangle +  \langle \phi_v^{(0)} | h_0 |\phi_v^{(1)} \rangle  \nonumber \\
&& +\sum_{a=1}^{N_c} \left [ \langle \phi_v^{(0)} \phi_a^{(0)} | g | \phi_v^{(1)} \phi_a^{(0)} \rangle - \langle \phi_v^{(0)} \phi_a^{(0)} | g | \phi_a^{(0)} \phi_v^{(1)} \rangle \right ] \nonumber \\
&&  +\sum_{a=1}^{N_c} \left [ \langle \phi_v^{(0)} \phi_a^{(0)} | g | \phi_v^{(0)} \phi_a^{(1)} \rangle - \langle \phi_i^{(0)} \phi_a^{(0)} | g | \phi_a^{(1)} \phi_v^{(0)} \rangle \right ] . \nonumber \\
\label{eqdfeng2}
\end{eqnarray}
Note that the core contributions are not included in the DHF expressions of the FF approach due to the reason mentioned earlier. It can be seen from the above expression that the DHF values of the IS factors in the FF approach contain extra terms  over the DHF expression of the EVE and AR approaches [see Eq.~(\ref{dhffn})]. They are the result of the modification of the single particle DHF orbital due to the presence of $O$ in the Hamiltonian of the FF approach. These extra contributions 
are known as orbital relaxation effects. They not only affect the DHF results, but can also contribute to the correlation effects in a many-body method.
However, they can be taken into account in the EVE and AR approaches by incorporating core-polarization effects (equivalent to RPA) due to the respective IS operator to all-orders in a many-body method.

Following the DHF procedure, we account for electron correlations either by the relativistic many-body perturbation theory (RMBPT) or, as described next, by the RCC method.
Although the RCC method is more potent in dealing with electron correlations, an RMBPT method helps to elucidate the magnitude of correlations and also to compare with prior calculations from the literature.

In RMBPT, the wave operator $\Omega_0$ is defined as
\begin{eqnarray}
\Omega_0 &=& 
\sum_{k=0}^{\infty} \Omega_0^{(k)} ,
\end{eqnarray}
where $\Omega_0^{(0)} =1$ and the superscript $k$ denotes the order of $V_\text{res}$. The amplitude of $\Omega_0^{(k)}$ can be determined using~\cite{Lindgren1985}
\begin{eqnarray}
\langle \Phi_0^* | [\Omega_0^{(k)},H_0] |\Phi_0 \rangle &=& \langle \Phi_0^* |V_\text{res} \Omega_0^{(k-1)} \nonumber \\
&& - \sum_{m=1}^{k-1} \Omega_0^{(k-m)} E_0^{(m-1)}]  |\Phi_0 \rangle ,
\end{eqnarray}
where $|\Phi_0^* \rangle $ denotes excited state Slater determinant with respect to $|\Phi_0\rangle$, and $E_0^{(m)}=\langle \Phi_0 | H \Omega_0^{(m-1)} |\Phi_0 \rangle$ is the total energy up to $m^{th}$-order corrections. Similarly, $\Omega_v$ can be expanded 
as
\begin{eqnarray}
\Omega_v &=& 
\sum_{k=0}^{\infty} \Omega_v^{(k)},
\end{eqnarray}
with $\Omega_v^{(0)} =1$. The amplitude-determining equation for the $\Omega_v^{(k)}$ wave operator is given by
\begin{eqnarray}
\langle \Phi_v^* | [\Omega_v^{(k)},H_0] |\Phi_v \rangle &=& \langle \Phi_v^* |V_\text{res} (\Omega_0^{(k-1)} + \Omega_v^{(k-1)}) \nonumber \\
&& - \sum_{m=1}^{k-1} \Omega_v^{(k-m)} E_v^{(m-1)}]  |\Phi_0 \rangle ,
\end{eqnarray}
where $|\Phi_v^* \rangle $ denotes the excited state Slater determinant with respect to $|\Phi_v\rangle$ and $E_v^{(m)}=\langle \Phi_0 | V_\text{res} (\Omega_0^{(m-1)} + \Omega_v^{(m-1)})|\Phi_0 \rangle$ is the energy of the valence state containing corrections up to $m^{th}$-order. The difference between $\sum_m E_v^{(m)}$ and $\sum_m E_0^{(m)}$ corresponds to the AE of the electron of the valence orbital $v$. 

In the presence of $O$ in the Hamiltonian, the wave operators can be redefined as
\begin{eqnarray}
\Omega_0^{(0)} = \sum_k \Omega_0^{(k,0)} \ \ \ \text{and} \ \ \  \Omega_v^{(0)} = \sum_k \Omega_v^{(k,0)}  
\end{eqnarray}
and
\begin{eqnarray}
\Omega_0^{(1)} = \sum_k \Omega_0^{(k,1)} \ \ \ \text{and} \ \ \  \Omega_v^{(1)} = \sum_k \Omega_v^{(k,1)},  
\end{eqnarray}
where the first and the second indices in the superscript of the wave operators denote the order of $V_\text{res}$ and $O$, respectively.

Since it deals with two different sources of perturbation simultaneously, it would still be difficult to carry out calculations up to a third-order perturbation in the RMBPT (RMBPT(3)) method. Thus, we only adopt the second-order RMBPT (RMBPT(2)) method approximation in the FF approach, which is equivalent to the RMBPT(3) method in the EVE and AR approaches. 

Accurate calculations of properties in neutral or singly charged systems, where electron correlations can be strong, require treating correlations non-perturbatively (“to all-orders”) using methods such as the RCC theory. 
The reason is that a truncated RCC theory is still an all-order method in which the wave functions and energies satisfy both size-consistent and size-extensivity behaviors.

In the RCC theory {\it Ansatz},
the wave operators are defined as~\cite{Lindgren1985,bijaya2,bijaya3}
\begin{eqnarray}
\Omega_0 = e^T \ \ \ \text{and} \ \ \ \Omega_v = e^T S_v .
\end{eqnarray}
The amplitude-determining equation of the core-orbital excitation operator $T$ is given by
\begin{eqnarray}
\langle \Phi_0^* |  \left ( H e^T \right )_l | \Phi_0 \rangle = 0 ,
\label{eqs0}
\end{eqnarray}
where $\bar{H}= (He^T)_l$ with the subscript $l$ denoting the linked terms. Similarly, the amplitude-determining equation for the valence-orbital excitation operator $S_v$ is given by
\begin{eqnarray}
 \langle \Phi_v^* | \{ (\bar{H}-E_v) S_v \} + \bar{H} | \Phi_v \rangle = 0 . \label{eqamp}
\end{eqnarray}

In the EVE and AR approaches, the unperturbed wave operators are denoted by
\begin{eqnarray}
\Omega_0^{(0)} = e^{T^{(0)}} \ \ \ \text{and} \ \ \ \Omega_v = e^{T^{(0)}} S_v{T^{(0)}} ,
\end{eqnarray}
which are obtained using the aforementioned RCC equations. Thus, the IS factors in the EVE approach are evaluated by 
\begin{eqnarray}
\langle O \rangle = \frac{\langle \Phi_v | \{1+S_v^{(0)} \}^{\dagger}  \bar{O} \{ 1+ S_v^{(0)} \} |\Phi_v \rangle} {\langle \Phi_v | \{1+ S_v^{(0)} \}^{\dagger} \bar{N} \{ 1+ S_v^{(0)} \} |\Phi_v \rangle} . 
\end{eqnarray}
As mentioned earlier, the above expression contains $\bar{O}=e^{T^{(0)\dagger}}Oe^{T^{(0)}}$ and $\bar{N}=e^{T^{(0)\dagger}}e^{T^{(0)}}$, which are two non-terminating terms.
To include as much contributions as possible from $\bar{O}$ and $\bar{N}$, we compute them by expressing sum of effective one-body, two-body and three-body terms. The dominant effective one-body terms are included self-consistently while effective two-body and three-body terms are computed directly.

The issue of non-terminating series in the IS expression of the EVE approach is taken care in the AR approach of RCC theory. Here, the the first-order perturbed wave operators are defined as~\cite{bijaya3,bijaya4,bijaya5}
\begin{eqnarray}
\Omega_0^{(1)} = e^{T^{(0)}} T^{(1)} 
\end{eqnarray}
and
\begin{eqnarray}
\Omega_v^{(1)} = e^{T^{(0)}} \left ( S_v^{(1)} + (1+S_v^{(0)})T^{(1)} \right )  .
\end{eqnarray}
The amplitude-determining equations for the first-order RCC operators are given by
\begin{eqnarray}
&& \langle \Phi_0^* |  \left ( H_\text{at} e^T T^{(1)} + O e^{T^{(0)}} \right )_l  | \Phi_0 \rangle = 0 \label{eqs01}\\
& \text{and} & \nonumber \\
&&  \langle \Phi_v^* | \left \{ \left ( H_\text{at} e^{T^{(0)}} \right )_l -E_v^{(0)})  \right \} S_v^{(1)}  + \left ( H_\text{at} e^{T^{(0)}} T^{(1)} \right )_l \nonumber \\
&&   \times  \left \{ 1+ S_v^{(0)} \right \} + \left ( O e^{T^{(0)}} \right )_l \left \{ 1+ S_v^{(0)} \right \} \nonumber \\
&& + \langle O \rangle S_v^{(0)} | \Phi_v \rangle = 0 ,  \ \ \ \label{eqv1} \ \ 
\label{eqamp1}
\end{eqnarray}
in which the IS is evaluated using the expression
\begin{eqnarray}
\langle O \rangle  &=& \langle \Phi_v | \left ( H_\text{at} e^{T^{(0)}} \right )_l S_v^{(1)}  + \left ( H_\text{at} e^{T^{(0)}} T^{(1)} \right )_l  \left \{ 1+ S_v^{(0)} \right \}   \nonumber \\
&& + ( O e^{T^{(0)}})_l \left \{1 + S_v^{(0)} \right \} | \Phi_v \rangle . 
\label{eqeng1}
\end{eqnarray}

\begin{table*}[tbb]
\caption{
Comparison of calculated AEs (in cm$^{-1}$) of the considered states in K with their respective experimental values from the NIST database~\cite{nist}. The experimental uncertainty is smaller than the last shown digit.
The estimated excitation energies (EEs) are also quoted from the AEs.
Our calculations are compared with similar ones from the literature at the relevant approximation level.
}
\begin{ruledtabular}
\begin{tabular}{l rrrr rrrr r}
State  & \multicolumn{1}{c}{DHF}  & \multicolumn{1}{c}{RMBPT(2)}  & \multicolumn{1}{c}{RCCSD} & \multicolumn{1}{c}{RCCSDT} & $+$Basis & $+$Breit  & $+$QED & \multicolumn{1}{c}{Total} & \multicolumn{1}{c}{Experiment}\\
\hline \\
AEs\\  
$4s~^2S_{1/2}$ & 32370.48  & 35077.13  & 35077.13 & 34972.90 & 16.47 & $-1.45$ & $-6.80$  & 34988(28) & $35009.81$  \\
~~Ref.~\cite{Eph94}  &     &           & 35028~ ~ ~&         &       &      ~ &        ~  & ~ ~ ~               \\
~~Ref.~\cite{Saf08}  & 32370~ ~ ~& 35104~ ~ ~& 34966~ ~ ~&         &       & $~1.7$~ & $-0.9$~  & 34967~ ~ ~               \\
~~Ref.~\cite{Yer23}  & 32373~ ~ ~& 35313~ ~ ~& 35138~ ~ ~&         &       &      ~ &        ~  & ~ ~ ~               \\
$4p~^2P_{1/2}$ & 21006.44  & 22010.82  & 22027.81 & 22018.00 & 8.30  & $-2.32$ & 0.47     & 22024(5)  & $22024.63$ \\
~~Ref.~\cite{Saf08}  & 21006~ ~ ~& 19993~ ~ ~& 22023~ ~ ~&         &       & $~1.3$~ & $0.0$~  & 22023  ~ ~               \\
~~Ref.~\cite{Yer23}  & 21000~ ~ ~& 22101~ ~ ~& 22079~ ~ ~&         &       &      ~ &        ~  & ~ ~ ~               \\
$4p~^2P_{3/2}$ & 20959.39  & 21951.66  & 21967.69 & 21957.90 & 8.26  & $-0.35$ & 0.18     & 21966(5)  & $21966.92$ \\
~~Ref.~\cite{Saf08}  & 20959~ ~ ~& 21960~ ~ ~& 21964~ ~ ~&         &       & $~0.3$~ & $0.0$~  & 21965 ~ ~               \\
~~Ref.~\cite{Yer23}  & 20960~ ~ ~& 22035~ ~ ~& 22013~ ~ ~&         &       &      ~ &        ~  & ~ ~ ~               \\
$5s~^2S_{1/2}$ & 13406.99  & 14028.06  & 13988.83 & 13979.84 & 3.63  & $-0.46$ & $ -0.16$ & 13983(3)  & 13983.26 \\
~~Ref.~\cite{Saf08}  & 13407~ ~ ~& 14035~ ~ ~& 13960~ ~ ~&         &       & $~0.3$~ & $-0.1$~  & 13960 ~ ~               \\
$5p~^2P_{1/2}$ & 10011.64  & 10313.36  & 10308.65 & 10307.84 & 2.57  & $-0.79$ & 0.16     & 10310(1)  & 10308.41 \\
~~Ref.~\cite{Saf08}  & 10012~ ~ ~& 10316~ ~ ~& 10304~ ~ ~&         &       & $-0.4$~ & $0.0$~  & 10304 ~ ~               \\
$5p~^2P_{3/2}$ & 9995.43   & 10293.90  & 10289.13 & 10288.24 & 2.55  & $-0.15$ & 0.07     & 10291(1)  & 10289.68 \\
~~Ref.~\cite{Saf08}  & 9996~ ~ ~&  10297~ ~ ~& 10285~ ~ ~&         &       & $~0.1$~ & $0.0$~  & 10285 ~ ~               \\
$6s~^2S_{1/2}$ & 7335.04   & 7574.11  &  7558.63 & 7555.61  & 1.42  & $-0.19$ & $-0.63$  & 7557(1)   & 7559.10 \\
~~Ref.~\cite{Saf08}  & 7338~ ~ ~& 7582~ ~ ~&   7548~ ~ ~&         &       & $~0.2$~ & $0.0$~  & 7549 ~ ~               \\
\vspace{1 mm}\\
EEs\\
$4s~^2S_{1/2}-4p~^2P_{1/2}$ & 11364.04 & 13066.31 & 13049.32 & 13015.90 & 8.17  & $0.87$   & $-7.27$ & 12964(24) & 12985.19 \\
~~Ref.~\cite{Eph94}  &     &           & 13012~ ~ ~&         &       &      ~ &        ~  & ~ ~ ~               \\
$4s~^2S_{1/2}-4p~^2P_{3/2}$ & 11411.09 & 13125.47 & 13109.44 & 13015.00 & 8.21  & $-1.10$  & $-6.98$ & 13022(24) & 13042.90 \\
$4s~^2S_{1/2}-5s~^2S_{1/2}$ & 18963.49 & 21049.07 & 21088.30 & 20993.06 & 12.84 & $-0.99$  & $-6.64$ & 21005(25) & 21026.55 \\
~~Ref.~\cite{Eph94}  &     &           & 21042~ ~ ~&         &       &      ~ &        ~  & ~ ~ ~               \\
$4s~^2S_{1/2}-5p~^2P_{1/2}$ & 22358.84 & 24763.77 & 24768.48 & 24665.06 & 13.90 &  $-0.66$ & $-6.96$ & 24678(27) & 24701.38 \\
~~Ref.~\cite{Eph94}  &     &           & 24722~ ~ ~&         &       &      ~ &        ~  & ~ ~ ~               \\
$4s~^2S_{1/2}-5p~^2P_{3/2}$ & 22375.05 & 24783.23 & 24788.00 & 24684.66 & 13.92 & $-1.30$  & $-6.87$ & 24697(27) & 24720.14 \\
$4s~^2S_{1/2}-6s~^2S_{1/2}$ & 25035.44 & 27503.02 & 27518.50 & 27417.29 & 15.05 & $-1.26$  & $-6.17$ & 27431(27) & 27450.71 \\
\end{tabular}
\end{ruledtabular}
\label{tab2}
\end{table*}

In essence, the commonality between the FF and AR methodologies lies in their utilization of energy-evaluating expressions, which are inherently truncated, for the assessment of IS factors.
However, there are two major differences between these two approaches. First, orbital relaxation effects are implicitly present in the FF approach while they appear only through RPA-type correlation effects in the AR approach. Second, it is presumed that the second- and higher-order contributions are negligibly small while estimating the first-order IS factors in the FF approach.
Both the EVE and AR methods employ unrelaxed atomic orbitals to directly calculate the first-order IS factors, disregarding higher-order contributions.  However, the EVE approach uses an expression that contains non-terminating series while the AR approach uses a naturally terminated expression. 

\section{Results and Discussion}

In this work, we first approximated the RCC theory using the RCCSD method to estimate contributions from the DC Hamiltonian, Breit interaction, and QED effects considering single-particle orbitals up to $g-$symmetry. To demonstrate the importance of contributions from orbitals with higher angular momentum, we repeat the calculations with the DC Hamiltonian after including orbitals from the $h-$, $i-$, and $j-$ symmetries. The differences between the RCCSD results from the two sets of orbitals are referred to as the ``$+$Basis" contributions. Another reason for performing calculations twice with the RCCSD method is that contributions from Breit and QED interactions from the higher-angular momentum orbitals are not significant, but it requires large computational resources to perform calculations including orbitals up to $j-$ symmetry. Calculations in the RCCSDT method are performed with orbitals up to $g-$symmetry. This is because both memory and computational time increase many-fold in the RCCSDT method compared to the RCCSD method, so it is impractical to obtain results for all the states of interest in K in the FF and AR approaches within a reasonable time with orbitals up to a higher symmetry. Lastly, we have also analyzed results in nonrelativistic and relativistic approximations with different nuclear charge distributions. 


\subsection{Attachment and Excitation Energies}

\begin{table}[tbp]
\caption{
Approximated nonrelativistic results of AE (in cm$^{-1}$) using the limit $c \rightarrow \infty$ in the DC Hamiltonian at different levels of atomic method. 
Our calculations are compared with other nonrelativistic calculations from the literature at the relevant approximation level.
$\Delta_\text{rel}$ is the difference between the relativistic and nonrelativistic values for $J=1/2$ states.
}
\begin{ruledtabular}
\begin{tabular}{l rrr}
State  & \multicolumn{1}{c}{DHF}  &  \multicolumn{1}{c}{RMBPT(2)} & \multicolumn{1}{c}{RCCSD} \\
\hline \\
$4S$ & 32252.78 & 34898.84 & 34892.58 \\
~~Ref~.\cite{Eph94} & &    & 34884 ~ ~ \\
~~$\Delta_\text{rel}$ & 117.70 & 178.29 & 184.55\\
$4P$ & 20971.47 & 21956.35 & 21986.00 \\
~~Ref~.\cite{Eph94} & &    & 21975 ~ ~ \\
~~$\Delta_\text{rel}$ & 34.97 & 54.47 & 41.81\\
$5S$ & 13375.77 & 13984.46 & 13953.57 \\
~~Ref~.\cite{Eph94} & &    & 13951 ~ ~ \\
~~$\Delta_\text{rel}$ & 31.22 & 43.60 & 35.26\\
$5P$ &  9999.54 & 10295.29 & 10294.88 \\
~~Ref~.\cite{Eph94} & &    & 10293 ~ ~ \\
~~$\Delta_\text{rel}$ & $-4.11$ & $-1.39$ & $-5.75$\\
$6S$ &  7323.13 &  7560.80 &  7545.50 \\
~~$\Delta_\text{rel}$ & $11.91$ & $13.31$ & $13.13$\\

\end{tabular}
\end{ruledtabular}
\label{tab1}
\end{table}

In Table~\ref{tab2}, we present the energies calculated with the DC Hamiltonian using the DHF, RMBPT(2), RCCSD and RCCSDT method approximations.
The final value quoted is the sum of contributions from the RCCSDT method, $+$Basis and $+$Breit. The approximate QED contribution is not added, but its full magnitude is considered an uncertainty.
The estimated uncertainty for the final value is based on the magnitudes of the triple contributions, basis extrapolation, and the approximate QED effect. 
Our results are consistent within this uncertainty with the experimental data~\cite{nist}, indicating that the wave functions from RCC theory can accurately determine the IS factors.

Compared to the studies in the literature~\cite{Eph94, Saf08, Yer23}, we observe strong alignment at the DHF level, suggesting that the basis size is sufficient.
At the RMBPT level, our findings are close to those of Safronova and Safronova~\cite{Saf08}, though they differ somewhat from Demidov \textit{et al.}~\cite{Yer23}. This could be due to slight variations in what is included in the RMBPT(2) definition. 
Our RCCSD approximation results match the linearized RCCSD calculations~\cite{Saf08,Yer23} within $2-3\times10^{-3}$.
The closest agreement is with the (nonlinearized) RCC results of Ref.~\cite{Eph94}. Although Breit and QED calculations differ mainly from those in Ref.~\cite{Saf08}, they are smaller than our final uncertainty. Nevertheless, further investigation is needed for more relativistic systems. To the best of our knowledge, earlier calculations did not consider triple excitations. Including triple excitations enhances agreement with the experiment by an order of magnitude for the ground level and all examined EEs.

In Table~\ref{tab1}, we show the nonrelativistic results for AEs when $c \rightarrow \infty$ using the DC Hamiltonian, highlighting the magnitude of relativistic effects on energies.
It is worth mentioning that these are not the exact nonrelativistic values as $c=1000$ a.u. is used in our calculations for practical purpose; which may still give some contributions from the smaller components of the Dirac orbitals.
Note that since identical computations are employed, the method names are kept the same as in the relativistic calculations.
Our data align well, within a few $10^{-4}$, with similar results from the literature~\cite{Eph94}.

When comparing the energies of a particular state from the methods in Tables~\ref{tab2} and~\ref{tab1}, it is evident that relativistic effects add approximately $0.4-0.5\%$ to the energies of the $4S$ state and $0.2-0.3\%$ for the $4P$, $5S$, and $6S$ states, thus critically influencing their proximity to experimental values. Nevertheless, the relativistic corrections are small enough that neglecting higher-order corrections beyond the Breit Hamiltonian for IS operators can be justified.

\subsection{Isotope shift factors using the FF approach}

With the RCCSDT method providing precise energy calculations through its wave functions, we proceeded to assess the IS factors using these wave functions. In this assessment, relativistic effects arise from the wave functions as well as from the expressions of the MS operators [Eq.~(\ref{nmsexp})]. In earlier calculations of MS factors, non-relativistic expressions for the operators were employed alongside relativistic wave functions. We repeat this calculation to compare with the values in the literature.
The results are given in Table~\ref{tab3} for different approximations in the many-body methods. For NMS factors, they show good agreement at both the DHF and RCCSD levels with a similar recent calculation~\cite{Sourav}.
Our SMS factors are in very good agreement with three other calculations at the DHF level~\cite{Sourav, Dzuba1, Safronova}.
At the MBPT level, our results align closely with those of Ref.~\cite{Dzuba1}, who also used the FF approach, while they diverge from those of Ref.~\cite{Safronova}, who relied on the EVE approach. Considerable differences were observed in SMS factors at the RCCSD approximation compared to those in~\cite{Sourav}, similar to the deviations observed for Mg$^+$~\cite{2022-Na}.

The electron correlation effects (deviations from the DHF values) in the determination of the SMS factors are stronger than for the NMS factors owing to the fact that the SMS operator is a two-body operator. The trends of the correlation effects for the SMS factors are such that the magnitudes become smaller in the RMBPT(2) method compared to the DHF results, then decrease further in the RCCSD method, then increase again in the RCCSDT method. They are also much stronger for the $4S$ (ground) state than for the other states. This suggests that missing higher-order correlation effects are not negligible, at least for the ground state. The magnitudes of the NMS factors from the DHF method are always smaller than the values obtained by employing a many-body method. The results of the RMBPT(2), RCCSD and RCCSDT methods are similar, suggesting that contributions from the higher-order correlation effects are negligible in this case.

\begin{table}[tb]
\caption{
Calculated NMS and SMS factors (in GHz amu) using their nonrelativistic expressions with the relativistic wavefunctions.
The FF approach is used. The results are compared with other calculations from the literature which have also employed nonrelativistic operators with relativistic wavefunctions.
$\Delta_\text{rel}$ is the difference between to the values obtained with the relativistic operators (given in Table.~\ref{tab:FF}).
}
\begin{ruledtabular}
\begin{tabular}{l rrrr}
State  & \multicolumn{1}{c}{DHF}  &  \multicolumn{1}{c}{RMBPT(2)} & \multicolumn{1}{c}{RCCSD} & \multicolumn{1}{c}{RCCSDT}\\
\hline \\   
\multicolumn{5}{c}{NMS factors}\\
$4S~^2S_{1/2}$ & 542.64 & 589.47 & 589.34 & 588.37 \\
~ Ref.~\cite{Sourav} & $549.4 ~$ & & $595.3$ ~\\
~ $\Delta_\text{rel}$& $-11.12$ & $-13.64$ & $-14.02$ & $-14.04$\\
$4p~^2P_{1/2}$ & 346.46 & 362.98 & 363.48 & 363.42 \\
~ Ref.~\cite{Sourav} & $354.8 ~$ & & $362.2$ ~\\
~ $\Delta_\text{rel}$& $-1.24$ & $-1.34$ & $-1.51$ & $-1.54$\\
$4p~^2P_{3/2}$ & 344.17 & 359.48 & 360.44 & 360.40 \\
~ Ref.~\cite{Sourav} & $348.5 ~$ & & $359.4$ ~\\
~ $\Delta_\text{rel}$& $0.68$ & $1.57$ & $0.92$ & $0.86$\\
$5s~^2S_{1/2}$ & 223.24 & 233.78 & 233.23 & 233.12 \\
~ $\Delta_\text{rel}$& $-3.03$ & $-3.41$ & $-3.45$ & $-3.45$\\
$5p~^2P_{1/2}$ & 165.03 & 169.97 & 169.96 & 169.97 \\
~ $\Delta_\text{rel}$& $-0.45$ & $-0.45$ & $-0.49$ & $-0.50$\\
$5p~^2P_{3/2}$ & 164.29 & 169.04 & 169.02 & 169.03 \\
~ $\Delta_\text{rel}$& $0.09$ & $0.21$ & $0.17$ & $0.17$\\
$6s~^2S_{1/2}$ & 121.89 & 126.25 & 125.94 & 125.90 \\
~ $\Delta_\text{rel}$& $-1.27$ & $-1.43$ & $-1.41$ & $-1.41$\\
     &          &          &         \\
\multicolumn{5}{c}{SMS factors}\\ 
$4S~^2S_{1/2}$ & $-195.75$ & $-54.89$  & $-12.51$ & $-29.46$\\
~ Ref.~\cite{Dzuba1}  & $-193.62$ & $-51.71$ 
\\
~ Ref.~\cite{Safronova}  & $-195.6$ ~& $-74.7~$ \\
~ Ref.~\cite{Sourav} & $-193.0$ ~& & $\mathit{-189.9}$\\
~ $\Delta_\text{rel}$& $-4.39$ & $-6.71$ & $-6.45$ & $-6.18$\\
$4p~^2P_{1/2}$ & $-59.60$  & $-23.84$  & $-11.87$ & $-19.65$ \\
~ Ref.~\cite{Dzuba1}  & $-61.51$ & $-25.95$
\\
~ Ref.~\cite{Safronova}  & $-59.6$ ~& $-24.7 ~$ \\
~ Ref.~\cite{Sourav} & $-66.1 ~$ & & $\mathit{-161.2}$\\
~ $\Delta_\text{rel}$& $2.55$ & $2.12$ & $2.78$ & $2.81$\\
$4p~^2P_{3/2}$ & $-58.83$  & $-23.26$  & $-11.56$ & $-19.19$ \\
~ Ref.~\cite{Safronova}  & $-59.0$ ~& $-24.3 ~$ \\
~ Ref.~\cite{Sourav} & $-58.1 ~$ & & $\mathit{-159.1}$\\
~ $\Delta_\text{rel}$& $0.06$ & $-0.41$ & $-0.38$ & $-0.33$\\
$5s~^2S_{1/2}$ & $-45.24$  & 0.14      & 6.53  & 6.24 \\
~ $\Delta_\text{rel}$& $-1.15$ & $-1.86$ & $-1.48$ & $-1.44$\\
$5p~^2P_{1/2}$ & $-20.85$  & $-7.45$   & $-4.75$ & $-5.95$ \\
~ $\Delta_\text{rel}$& $0.96$ & $0.98$ & $1.00$ & $1.01$\\
$5p~^2P_{3/2}$ & $-20.60$  & $-7.28$   & $-4.64$ & $-5.89$ \\
~ $\Delta_\text{rel}$& $0.08$ & $-0.06$ & $-0.05$ & $0.04$\\
$6s~^2S_{1/2}$ & $-17.66$  & 1.44      & 3.56 & 4.07 \\
~ $\Delta_\text{rel}$& $-0.41$ & $-0.59$ & $-0.53$ & $0.70$\\
\end{tabular}
\end{ruledtabular}
\label{tab3}
\end{table}

\begin{table*}[tbp]
\caption{
Calculated IS factors at different levels of approximation in the FF approach.
The last column indicates the semi-empirical value obtained by scaling the experimental energies by the electron mass in amu. Its uncertainty is estimated by scaling the relativistic corrections to the energies ($\Delta_\text{rel}$ of Table~\ref{tab1}).
}
\begin{ruledtabular}
\begin{tabular}{l rrrr rrrr c}
    & \multicolumn{1}{c}{DHF}  & \multicolumn{1}{c}{RMBPT(2)}  & \multicolumn{1}{c}{RCCSD} & \multicolumn{1}{c}{RCCSDT} & $+$Basis & $+$Breit  & $+$QED & \multicolumn{1}{c}{Total} & \multicolumn{1}{c}{Scaling~\cite{nist}} \\
\hline \\

\multicolumn{10}{c}{$K^\text{NMS}$ values (in GHz amu)} \\
$4s~^2S_{1/2}$ & 531.52 & 575.83 & 575.32 & 574.33 & 0.24 & $-0.08$ & $-0.29$ & 574.5(5) & 575.8(3.0) \\
$4p~^2P_{1/2}$ & 345.22 & 361.64 & 361.97 & 361.88 & 0.12 & $-0.53$ & $-0.09$ & 361.5(3) & 362.2(7) ~ \\
$4p~^2P_{3/2}$ & 344.85 & 361.05 & 361.36 & 361.26 & 0.12 & $-0.51$ & $-0.52$ & 360.9(6) & 361.3(7) ~ \\
$5s~^2S_{1/2}$ & 220.21 & 230.37 & 229.78 & 229.67 & 0.05 & $-0.03$ & $-0.09$ & 229.7(1) & 223.0(6) ~ \\
$5p~^2P_{1/2}$ & 164.58 & 169.52 & 169.47 & 169.47 & 0.03 & $-0.09$ & $-0.06$ & 169.4(1) & 169.5(1) ~ \\
$5p~^2P_{3/2}$ & 164.38 & 169.25 & 169.19 & 169.20 & 0.04 & $-0.05$ & $-0.13$ & 169.2(1) & 169.2(1) ~ \\
$6s~^2S_{1/2}$ & 120.62 & 124.82 & 124.53 & 124.49 & 0.06 & $-0.12$ & $-0.15$ & 124.4(2) & 124.3(2) ~ \\
\\

$4s$$~^2S_{1/2}$$-4p~^2P_{1/2}$   & 186.30 & 214.19 & 213.35 & 212.45 & 0.12 & $ 0.45$ & $-0.20$ & 213.0(4) & 213.6(2.3) \\
$4s$$~^2S_{1/2}$$-4p~^2P_{3/2}$   & 186.67 & 214.78 & 213.96 & 213.07 & 0.12 & $ 0.43$ & $ 0.23$ & 213.6(4) & 214.5(2.3) \\
$4s$$~^2S_{1/2}$$-5s$$~^2S_{1/2}$ & 311.31 & 345.46 & 345.54 & 344.66 & 0.19 & $-0.05$ & $-0.20$ & 344.8(4) & 345.8(2.5) \\
$4s$$~^2S_{1/2}$$-5p~^2P_{1/2}$   & 366.94 & 406.31 & 405.85 & 404.86 & 0.21 & $ 0.01$ & $-0.23$ & 405.1(4) & 406.2(3.1) \\
$4s$$~^2S_{1/2}$$-5p$$~^2P_{3/2}$ & 367.14 & 406.58 & 406.13 & 405.13 & 0.20 & $-0.03$ & $-0.16$ & 405.3(4) & 406.6(3.1) \\
$4s$$~^2S_{1/2}$$-6s$$~^2S_{1/2}$ & 410.90 & 451.01 & 450.79 & 449.84 & 0.18 & $ 0.04$ & $-0.14$ & 450.1(4) & 451.5(2.8) \\\\

\multicolumn{10}{c}{$K^\text{SMS}$ values (in GHz amu)} \\
$4s~^2S_{1/2}$   & $-200.14$& $-61.60$ & $-18.96$& $-35.64$ & $-0.45$ & 0.56    & 0.27   & $-36(6)$\\
$4p~^2P_{1/2}$   & $-57.05$ & $-21.14$ & $-9.09$ & $-16.84$ & $-0.08$ & 0.07    & 0.25   & $-17(3)$\\
$4p~^2P_{3/2}$   & $-58.77$ & $-23.67$ & $-11.94$& $-19.52$ & $-0.08$ & 0.48    & 0.25   & $-19(3)$\\
$5s$$~^2S_{1/2}$ & $-46.39$ & $-1.72$  & 5.05    & 4.80     & $-0.13$ & 0.04    & $-0.06$& $4.7(1)$  \\
$5p~^2P_{1/2}$   & $-19.89$ & $-6.47$  & $-3.75$ & $-4.94$  & $-0.03$ & $-0.01$ & $-0.09$& $-5.0(4)$ \\
$5p~^2P_{3/2}$   & $-20.52$ & $-7.34$  & $-4.69$ &  $-5.85$ & $-0.03$ & 0.14    & 0.02   & $-5.7(4)$ \\
$6s$$~^2S_{1/2}$ & $-18.07$ & 0.88     & 3.03    & 3.37     & $-0.07$ & 0.03    & 0.03   & $3.3(1)$  \\
\\
$4s$$~^2S_{1/2}$$-4p~^2P_{1/2}$  &$-143.09$ &$-40.46$& $-9.87$ & $-18.8$  & $-0.37$ & $0.49$&$0.02$&$-18.7(3.0)$\\
$4s$$~^2S_{1/2}$$-4p~^2P_{3/2}$  &$-140.54$ &$-37.93$& $-7.02$ & $-16.12$ & $ -0.37$& $0.08$&$0.02$&$-16.4(3.0)$\\
$4s$$~^2S_{1/2}$$-5s$$~^2S_{1/2}$&$-153.75$ &$-59.88$& $-24.01$& $ -40.44$& $-0.32$ & $0.52$&$0.33$&$-40.2(5.5)$\\
$4s$$~^2S_{1/2}$$-5p~^2P_{1/2}$  &$-180.25$ &$-55.13$& $-15.21$& $-30.7$  & $-0.42$ & $0.57$&$0.36$&$-30.6(5.2)$\\
$4s$$~^2S_{1/2}$$-5p$$~^2P_{3/2}$&$-179.62$ &$-54.26$& $-14.27$& $-29.79$ & $-0.42$ & $0.42$&$0.25$&$-29.8(5.2)$\\
$4s$$~^2S_{1/2}$$-6s$$~^2S_{1/2}$&$-182.07$ &$-60.72$& $-21.99$& $-39.01$ & $-0.38$ & $0.53$&$0.24$&$-38.9(5.7)$\\\\

\multicolumn{10}{c}{$F$ values (in MHz/fm$^{-2}$) } \\
$4s~^2S_{1/2}$ & $-80.25$ & $-106.91$ & $-106.16$ & $-105.50$ & $-0.14$   & 0.13      &  1.24    & $-105.5(1.3)$ \\
$4p~^2P_{1/2}$ & 5.05     & 4.38      & 4.65      & 4.55      & $\sim0.0$ & $\sim 0.0$& $-0.04$  & 4.55(5) \\
$4p~^2P_{3/2}$ & 5.17     & 4.58      & 4.85      & 4.75      & $\sim0.0$ & $-0.01$   & $-0.04$  & 4.75(5) \\
$5s~^2S_{1/2}$ & $-21.20$ & $-25.83$  & $-25.32$  & $-25.25$  & $-0.02$   & 0.03      & 0.29     & $-25.24(29)$ \\
$5p~^2P_{1/2}$ & 1.92     & 1.82      & 1.91      & 1.88      & $\sim0.0$ & $-0.01$   & $-0.02$  & 1.87(2) \\
$5p~^2P_{3/2}$ & 1.88     & 1.78      & 1.86      & 1.84      & $\sim0.0$ & $\sim 0.0$& $-0.01$  & 1.83(1) \\
$6s~^2S_{1/2}$ & $-8.54$  & $-10.65$  & $-10.40$  & $-10.35$  & $-0.03$   & $0.01$    & 0.11     & $-10.37(11)$ \\
\\

$4s$$~^2S_{1/2}$$-4p~^2P_{1/2}$ & $-85.30$& $-111.29$ & $-110.81$ & $-110.05$ & $-0.14$ & $0.13$ &$1.28$ &$-110.1(1.3)$ \\
$4s$$~^2S_{1/2}$$-4p~^2P_{3/2}$   & $-85.42$  & $-111.49$ & $-111.01$ & $-110.25$ & $-0.14$ & 0.14 & 1.28 & $-110.3(1.3)$\\
$4s$$~^2S_{1/2}$$-5s$$~^2S_{1/2}$ & $-59.05$ & $-81.08$ & $-80.84$ & $-80.25$ & $-0.12$ & 0.11 & 0.95     &  $-80.3(1.0)$\\
$4s$$~^2S_{1/2}$$-5p~^2P_{1/2}$   & $-82.17$  & $-108.73$ & $-108.07$ & $-107.38$ & $-0.14$ & 0.14 & 1.26 &  $-107.4(1.3)$\\
$4s$$~^2S_{1/2}$$-5p~^2P_{3/2}$   & $-82.13$  & $-108.69$ & $-108.02$ & $-107.34$ & $-0.14$ & 0.13 & 1.27 & $-107.3(1.3)$ \\
$4s$$~^2S_{1/2}$$-6s$$~^2S_{1/2}$ & $-71.71$ & $-96.26$ & $-95.76$ & $-95.15$ & $-0.11$ & 0.12 & 1.13     & $-95.1(1.1)$ 

\end{tabular}
\end{ruledtabular}
\label{tab:FF}
\end{table*}

In Table~\ref{tab:FF}, we present the results of the IS factor calculations using relativistic wave functions and relativistic expressions of the MS operators [Eq.~(\ref{nmsexp})]. We compare these values with the values obtained using the non-relativistic operators.
The differences between the values, whose magnitude is a few percent, are presented in Table~\ref{tab3} for each state and many-body method. It is observed that most of the difference occurs at the DHF level, with minor adjustments attributable to electron correlations. It is also most prominent for the ground state, as expected. Additionally, we notice that, while they are significant at the current level of accuracy, their magnitude suggests that higher-order relativistic corrections to the operators are insignificant.

The calculated NMS factors are very close to their values obtained by scaling the experimental energies from the NIST database~\cite{nist}. As the scaling law is precise in the non-relativistic limit, it suggests that using relativistic wave functions with non-relativistic operators may overestimate the relativistic contributions. It is thus clear that relativistic operators must be used with relativistic wave functions.
The nearly identical results obtained with the RMBPT(2), RCCSD, and RCCSDT methods suggest that the higher-level electron correlations are small, which makes the calculated NMS factors very accurate. 

The results of the SMS factors calculated using the relativistic operator are also given in Table~\ref{tab:FF}. Their magnitudes drastically reduce going from the DHF values to RMBPT(2) and then RCCSD, but increase again at the RCCSDT approximation. These trends are similar to those of the non-relativistic operator that are given in Tables~\ref{tab3}. The large differences between the RCCSD and RCCSDT results imply that contributions from the quadruple excitations may not be small. In fact, they are presumed to be larger than the $+$Basis, $+$Breit and $+$QED corrections. We estimate the corresponding uncertainty as a third of the difference between the RCCSD and RCCSDT values.
This fits with the accuracy of the excitation energies as well as the experimental benchmarks given later.

\begin{table*}[tbp]
\caption{IS factors and $A_{hf}$ values of $^{39}$K using $g_I=0.261005$ from the DHF and RCC methods in the EVE approach.}
\begin{ruledtabular}
\begin{tabular}{l rrr rrrr r}
State  & \multicolumn{1}{c}{DHF}  & \multicolumn{1}{c}{RCCSD} & \multicolumn{1}{c}{RCCSDT} & $+$Basis & $+$Breit  & $+$QED & \multicolumn{1}{c}{Total} & \multicolumn{1}{c}{Experiment}\\
\hline \\

\multicolumn{9}{c}{$K^\text{NMS}$ values (in GHz amu)} \\
$4s~^2S_{1/2}$ & 941.13 & 594.22 & 553.75 & 1.12 & $-0.44$ & $-0.38$    & 554\\
$4p~^2P_{1/2}$ & 488.53 & 357.97 & 339.14 & 0.49 & $-0.08$ & 0.04       & 340\\
$4p~^2P_{3/2}$ & 486.87 & 356.79 & 338.19 & 0.48 & 0.01 & 0.02          & 339\\
$5s~^2S_{1/2}$ & 324.53 & 235.11 & 224.69 & 0.43 & 0.09 & $-0.09$       & 225\\
$5p~^2P_{1/2}$ & 214.14 & 169.33 & 162.88 &  0.13 & $-0.03$ & 0.01      & 163\\
$5p~^2P_{3/2}$ & 213.61 & 168.98 & 162.63 & 0.13 & $-0.01$ & $\sim 0.0$ & 163\\
$6s~^2S_{1/2}$ & 162.52 & 126.73 & 122.64 & 0.09 & $\sim 0.0$ & $-0.03$ & 123\\\\

\multicolumn{9}{c}{$K^\text{SMS}$ values (in GHz amu)} \\
$4s~^2S_{1/2}$ & $-388.76$ & $-57.87$ & $-106.84$ & 0.22 & 0.21 & 0.37 & $-106$ & \\
$4p~^2P_{1/2}$ & $-115.54$ & $-21.44$ & $-42.02$ & 0.32 & $\sim 0.0$ & $-0.06$ & $-42$ &  \\
$4p~^2P_{3/2}$ & $-116.33$ & $-20.52$ & $-41.05$ & 0.31 & $-0.02$ & $-0.05$ & $-41$ & \\
$5s~^2S_{1/2}$ & $-94.60$ & $-4.90$ & $-11.42$ & 0.02 & 0.15 & 0.07 & $-11$ & \\
$5p~^2P_{1/2}$ & $-40.19$ & $-8.68$ & $-12.93$ & 0.10 & 0.01 & $-0.02$ & $-13$ & \\
$5p~^2P_{3/2}$ & $-40.49$ & $-8.36$ & $-12.63$ & 0.10 & 0.02 & $-0.01$ & $-13$ & \\
$6s~^2S_{1/2}$ & $-37.49$ & $-0.99$ & $-2.97$ & 0.01 & 0.06 & 0.03 & $-3$ & \\\\

\multicolumn{9}{c}{$F$ values (in MHz/fm$^{-2}$) } \\
$4s~^2S_{1/2}$ & $-73.08$ & $-103.95$ & $-104.01$ & $-0.09$ & 0.22 & 1.23 & $-103.9(1.2)$ & $-103.59...$~\cite{Arimondo}\\
~ Ref.~\cite{Martensson2}  & & $-103$ ~ ~ &  \\
~ Ref.~\cite{Safronova}  & $-73.03$ & $-106.72$ &  \\
~ Ref.~\cite{Dzuba1}  &&  \multicolumn{2}{c}{$-104.20$}\\
$4p~^2P_{1/2}$ & $-0.08$ & 4.48 & 4.70 & 0.01 & $-0.01$ & $-0.05$ & $4.7(1)$ & \\
~ Ref.~\cite{Martensson2}  & & $4.6$~ &  \\
~ Ref.~\cite{Safronova}  & $-0.08$ & $3.81$ &  \\
~ Ref.~\cite{Dzuba1}  &&  \multicolumn{2}{c}{~ ~ $4.04$}\\
$4p~^2P_{3/2}$ & $\sim0.0$ & 4.54 & 4.75 & 0.01 & $-0.01$ & $-0.05$ & $4.8(1)$ & \\
~ Ref.~\cite{Safronova}  & $0.00$ & $3.90$ &  \\
$5s~^2S_{1/2}$ & $-19.34$ & $-24.48$ & $-24.53$ & $-0.01$ & $\sim 0.0$ & 0.28 & $-24.5(3)$ & $-24.9(3)$~\cite{Arimondo}\\
$5p~^2P_{1/2}$ & $-0.03$ & 1.56 & 1.64 & $\sim 0.0$ & $\sim 0.0$ & $-0.02$ & 1.64(3) & \\
$5p~^2P_{3/2}$ & $\sim 0.0$ & 1.58 & 1.66 & $\sim 0.0$ & $\sim 0.0$ & $-0.02$ & 1.66(3) & \\
$6s~^2S_{1/2}$ & $-7.86$ & $-9.64$  & $-9.66$ & $-0.01$ & $\sim 0.0$ & 0.11 & $-9.67(11)$ & $-9.79(8)$~\cite{2022-HFSrev}\\\\

\multicolumn{9}{c}{$A_{hf}$ values (in MHz) of $^{39}$K} \\
$4s~^2S_{1/2}$ & 146.91 & 230.68 & 230.38 & 0.34 & $\sim 0.0$  & $-1.42$ & 230.7(1.4) & 230.86...~\cite{Arimondo} \\
$4p~^2P_{1/2}$ & 16.62 & 27.32 & 27.56 & $\sim 0.0$ & 0.01 & $\sim 0.0$ & 27.6(1) ~ & 27.79(7)~\cite{2022-HFSrev} \\
$4p~^2P_{3/2}$ & 3.23 & 5.97 & 5.97 & $\sim 0.0$ & $\sim 0.0$ & $-0.01$ & 5.97(5) & 6.08(2)~\cite{2022-HFSrev} \\
$5s~^2S_{1/2}$ & 38.88 & 55.09 & 55.20 & 0.06 & 0.11 & $-0.33$ & 55.4(3) & 55.5(6)~\cite{Arimondo}\\
$5p~^2P_{1/2}$ & 5.74 & 8.85 & 8.93 & $\sim0.0$ & 0.03 & $-0.01$ & 8.96(5) & 9.01(17)~\cite{2022-HFSrev} \\
$5p~^2P_{3/2}$ & 1.12 & 1.93 & 1.94 & $\sim 0.0$ & $\sim 0.0$ & $\sim 0.0$ & 1.94(5) & 1.97(1)~\cite{Arimondo} \\
$6s~^2S_{1/2}$ & 15.79 & 21.78 & 21.84 & 0.02 & 0.02 & $-0.13$ & 21.9(1) & 21.89(9)~\cite{2022-HFSrev} \\
\end{tabular}
\end{ruledtabular}
\label{tab7}
\end{table*}

The FS factors are given in Table~\ref{tab:FF}.
Their relative magnitudes are as expected from a single-valence system, with contributions to the $S$ states much larger than to the $P$ states. The correlation trends are also different. For the $S$ states, $F$ increases from the DHF values, while this trend is opposite in the $P$ states. Like the NMS factor, the differences among RMBPT(2), RCCSD, and RCCSDT are below 1\%, suggesting that the electron correlations are small and have been well taken into account. However, unlike as with the MS factors, the QED contributions are not negligible, owing to its ability to alter the atomic wave-functions in the vicinity of the nucleus.
Similarly to the energies, we consider the complete QED contribution of approximately 1\% as an uncertainty. Although it is the main source of uncertainty for $F$, it remains insignificant overall due to the uncertainty in $K^\text{SMS}$.

Up to this point, we have explored the findings for IS factors derived using the FF method. The uncertainties reported for the NMS and FS factors are significantly smaller than those for the SMS factors. Our capacity to accurately estimate the uncertainties due to various numerical integrations and differentiations involved throughout these calculations is somewhat constrained. Furthermore, the second-order contributions inherent in the FF method are typically considered negligible. This assumption might hold for light elements such as K, though it could differ for heavier elements. Furthermore, accurate assessments of second-order effects on IS factors are important for evaluating nonlinear contributions to King plots, which could then be distinguished from new physics~\cite{Berengut2}.
Consequently, it would be beneficial to validate the findings from the EVE and AR methods mentioned above against the FF method. The results of these approaches are elaborated on in the following.

\subsection{IS and hyperfine factors from the EVE approach}

In Table~\ref{tab7}, we present the IS factors calculated with the EVE approach at different levels of approximation. Relativistic wave functions and MS operators are employed.
As can be seen in the table, the DHF values of the FS, NMS, and SMS factors are different from their corresponding values in Table~\ref{tab:FF}.
This is due to the orbital relaxation effects present when the DHF method is used with the FF approach and absent when the EVE or AR approaches are used. 
This makes the DHF results with the FF approach much closer to their \text{real} values, which include electron correlations.
In the EVE and AR approaches, which use unrelaxed orbitals,
the RPA contributions embedded in the RCC theory should account for these differences.
However, it is not guaranteed that the relaxation effects of all occupied and unoccupied orbitals of the FF approach can be included in the RCCSD or RCCSDT method approximation in the EVE approach. 

Another notable difference is that the EVE approach does not obey the size-extensivity behavior, in contrast to the FF approach. Also, both the numerators and denominators of the EVE approach contain non-terminating series so that we have terminated the expressions forcefully~\cite{bijaya2,bijaya3}.
Specifically for the SMS factors, some of the forcefully neglected terms correspond to at least a third-order perturbation in the FF approach. 
As a result, we cannot ensure how much the higher-order terms of the EVE expression would contribute in our calculations, and so we could not quote a reasonable uncertainty for the MS factors. 
Alternative procedures can be used to estimate uncertainties in the IS factors of the EVE approach, but they are not pursued in this work.

Considering that the $+$Basis, $+$Breit and $+$QED corrections estimated in the EVE approach are also found to be small and that the NMS values are far from the scaling-law values,
The large differences between the final results of both the FF and the EVE approaches cannot be justified.
We surmise that the MS factor calculations in K using the EVE approach suffer from unquantified numerical uncertainties. 
This explains the differences observed between the MS factors of this work and other works using the EVE approach~\cite{Safronova, Sourav}, and the agreement with the work employing the FF approach~\cite{Dzuba1}.

In contrast to the MS factors, the FS factors calculated with the EVE approach agree reasonably well, considering the uncertainty of the QED contributions, with those calculated with the FF approach. In light of this, agreement is achieved with all other calculations in the literature.
Further empirical testing of the FS factor using a calibrated King Plot is not possible as the radii of only two isotopes of K have been measured with muonic atom x-ray spectroscopy~\cite{1981-WSH}. Thus, we pursue a different method. First, the $A_{hf}$ values are calculated using the EVE approach and presented in Table~\ref{tab7} in a similar manner to the IS factors. 
Our final values, in which QED contributions are not included but taken as uncertainty, are in good agreement with the measured values, to within $1\%$.
These results show that the electron correlation and QED effects in the FS factors and the $A_{hf}$ values follow similar trends, which supports our uncertainty estimation of the FS factors.
Furthermore, for the $S$ states, we verified the simple relation between the FS factors and the $A_{hf}$ values of Eq.~(\ref{eqFA}), within the combined uncertainties.

\begin{table*}[tbp]
\caption{IS factors from the AR approach using DHF and RCC methods.
Our results for the $D_2$ transitions are compared with the previous calculation which also utilized the AR method.}
\begin{ruledtabular}
\begin{tabular}{l rrr rrrr }
State  & \multicolumn{1}{c}{DHF}  & \multicolumn{1}{c}{RCCSD} & \multicolumn{1}{c}{RCCSDT} & $+$Basis & $+$Breit  & $+$QED & \multicolumn{1}{c}{Total} \\
\hline \\

\multicolumn{8}{c}{$K^\text{NMS}$ values (in GHz amu)} \\
$4s~^2S_{1/2}$ & 941.13 & 559.22 & 545.45 & $-1.12$ & $-1.79$ & $-2.06$ & 543(5)\\
$4p~^2P_{1/2}$ & 488.53 & 351.99 & 345.57 & $-1.45$ & $-1.81$ & $-1.69$ & 342(3)\\
$4p~^2P_{3/2}$ & 488.87 & 351.69 & 345.41 & $-1.17$ & $-1.43$ & $-1.40$ & 343(3)\\
$5s~^2S_{1/2}$ & 324.53 & 226.78 & 222.47 & 0.37 & 0.21 & 0.15 & 223(1)\\
$5p~^2P_{1/2}$ & 214.14 & 166.99 & 164.33 & $-0.24$  & $-0.35$ & $-0.31$ & 164(1)\\
$5p~^2P_{3/2}$ & 213.61 & 166.88 & 164.23 & $-0.15$ & $-0.23$ & $-0.21$ & 164(1)\\
$6s~^2S_{1/2}$ & 162.52 & 123.50 & 121.20 & 0.04 & $-0.02$ & $-0.04$ & 121(1)\\\\
\multicolumn{8}{c}{$K^\text{SMS}$ values (in GHz amu)} \\
$4s~^2S_{1/2}$ & $-388.76$ & $-12.72$  & $-25.05$ & $-0.53$ & 0.48 & 0.31 & $-25(4)$\\
$4p~^2P_{1/2}$ & $-115.54$ & $-5.34$ & $-5.33$  & $-0.14$  & 0.22 & $-0.06$ & $-5.3(1)$\\
$4p~^2P_{3/2}$ & $-116.33$ & $-8.95$ & $-9.48$ & $-0.13$ & 0.11 & $-0.06$ & $-9.5(2)$\\
$5s~^2S_{1/2}$ & $-94.60$ & 6.05 & 9.04 & $-0.15$ & 0.13 & 0.06 & 9.0(1)\\
$5p~^2P_{1/2}$ & $-40.19$ & $-2.98$ & $-0.33$ & $-0.07$ &  0.08 & $-0.03$ & $-0.3(9)$\\
$5p~^2P_{3/2}$ & $-40.49$ & $-4.13$ & $-1.78$ & $-0.05$ & 0.04 & $-0.15$ & $-1.8(8)$\\
$6s~^2S_{1/2}$ & $-37.49$ & 3.34 & 5.80 & $-0.07$ & 0.06 & 0.02 & 5.8(8)\\
\\
$4s$$~^2S_{1/2}$$-4p~^2P_{1/2}$ & $-273.22$& $-7.38$ & $-19.72$ &$-0.38$ & $0.26$ & $0.37$ & $-20(4)$ ~ ~ \\
~ ~Ref.~\cite{Koszorus} &&&&&&& $-14.0(2.2)$ \\\\

\multicolumn{8}{c}{$F$ values (in MHz/fm$^{-1}$)} \\
$4s~^2S_{1/2}$ & $-73.08$ & $-105.20$ & $-103.62$ & $-0.12$ & 0.13  & 1.24 & $-103.6(1.4)$ \\
$4p~^2P_{1/2}$ & $-0.08$  & 4.05 & 3.80 & $0.01$ & $\sim 0.0$ & $-0.05$ & 3.8(1)\\
$4p~^2P_{3/2}$ & $\sim 0.0$ & 4.17 & 3.87 & $0.01$ & $-0.01$ & $-0.04$ & 3.9(1)\\
$5s~^2S_{1/2}$ & $-19.34$ & $-24.76$ & $-24.53$ & $-0.02$ & 0.03 & 0.29 & $-24.5(3)$\\
$5p~^2P_{1/2}$ & $-0.03$ & 1.45 & 1.35 & $\sim 0.0$ & $-0.01$ & $-0.02$ & 1.34(4)\\
$5p~^2P_{3/2}$ & $\sim 0.0$ & 1.48 & 1.37 & $0.01$ & $\sim 0.0$ & $-0.01$ & 1.38(4)\\
$6s~^2S_{1/2}$ & $-7.86$ & $-9.73$ & $-9.67$ & $-0.01$ & $0.01$ & 0.11 & $-9.67(11)$\\\\
$4s$$~^2S_{1/2}$$-4p~^2P_{1/2}$ & $-73.00$& $-109.25$ & $-107.42$ & $-0.13$ & $0.13$ & $1.29$ &$-107.4(1.4)$  \\
~ ~Ref.~\cite{Koszorus} &&&&&&& $-107.2(5)$\hspace{7 pt} \\
\end{tabular}
\end{ruledtabular}
\label{tab8}
\end{table*}

\subsection{IS factors with the AR approach}

The large deviations of the MS factors of the EVE approach compared with the FF approach are understood to be mainly due to the orbital relaxation effects and non-terminating terms of the EVE approach in the RCC theory.
Thus, it would be interesting to repeat the calculations with the AR approach, which does not contain non-terminating series, like the FF approach, but in which orbital relaxation effects are absent, as with the EVE approach.
The AR results for the IS factors are given in Table~\ref{tab8}. 
Comparing $K^\text{SMS}$ for the $D1$ line with that given in Ref.~\cite{Koszorus} we find a slight deviation compared to the combined uncertainty. This is ascribed to the larger basis set used here. Our basis extrapolation shift ensures that the current basis set is large enough.

As can be seen in this table, the DHF values of the IS factors in this approach are the same as those of the EVE approach; however, the final values are very different. For the MS factors, the $+$Basis, $+$Breit and $+$QED contributions estimated in this approach are also different from the two approaches discussed previously. For the FS factors, they are similar.

The electron correlation patterns of IS factors exhibited in the AR approach align more closely with those observed in the EVE approach compared to the FF approach, despite significant disparities in their magnitudes.
The NMS factors observed across the three methodologies differ, with only the FF approach approximating the scaling law values, which are expected to be accurate, at least for the higher lying states in which the relativistic effects are smaller.

\begin{table*}[tbp]
\caption{
Calculated FS factors (in MHz/fm$^2$) using different charge distribution and density of electron at the origin approximations using DC Hamiltonian in the AR approach of the DHF and RCCSD method. 
}
\begin{ruledtabular}
\begin{tabular}{l rrrr rrrr}
State  & \multicolumn{4}{c}{ DHF}  & \multicolumn{4}{c}{RCCSD}  \\
\cline{2-5} \cline{6-9} \\
 & $\rho(0)$ & Uniform & Gaussian &  Fermi & $\rho(0)$ & Uniform & Gaussian &  Fermi \\
\hline \\
$4s~^2S_{1/2}$ & $-73.44$ & $-73.15$  & $-73.00$  & $-73.08$  &  $-105.69$ & $-105.26$ &  $-105.04$ & $-105.20$ \\
$4p~^2P_{1/2}$ & $-0.08$ & $-0.08$  & $-0.08$  & $-0.08$  &  $4.06$  & 4.04 & 4.04 & 4.05 \\
$4p~^2P_{3/2}$ & $\sim 0.0$ & $\sim 0.0$  & $\sim 0.0$  &  $\sim 0.0$ & 4.17  &  4.16 & 4.15 & 4.17 \\
$5s~^2S_{1/2}$ & $-19.44$ & $-19.36$ & $-19.32$  & $-19.34$  &  $-24.88$  & $-24.78$ & $-24.73$ & $-24.76$ \\
$5p~^2P_{1/2}$ & $-0.03$ & $-0.03$  & $-0.03$  & $-0.03$  &  1.45  &  1.44 & 1.44 & 1.45 \\
$5p~^2P_{3/2}$ & $\sim 0.0$ & $\sim 0.0$  & $\sim 0.0$  & $\sim 0.0$   &  1.49  & 1.48 & 1.48 &  1.48 \\
$6s~^2S_{1/2}$ & $-7.90$ & $-7.86$  & $-7.85$  & $-7.86$  & $-9.78$ &  $-9.74$  &  $-9.72$ & $-9.73$ \\
\end{tabular}
\end{ruledtabular}
\label{tab9}
\end{table*}

The SMS factors for the states derived in the three approaches vary considerably; however, the differential values for the transitions derived from the AR and FF approaches agree within $1-2$ times their combined uncertainty. 
As the correlation trends of both methods are rather different, this gives confidence in the values obtained. 
The deviations observed between the FF approach and both the EVE and AR approaches may not be attributable to orbital relaxation but potentially to unidentified physical phenomena, which warrants further investigation.

All of the discussed IS factors were determined using the Fermi nuclear charge distribution.
Our prior work on similar systems suggests that, in a light system such as K,  model dependence could affect the FS factor, but should be negligible compared with other sources of uncertainty~\cite{2024-Ag}.
To verify this, we used four different expressions for the FS operator, as discussed in Ref.~\cite{review}, to verify the reliability of the calculations of $F$. First, an expression that involves the electron density at the center of the nucleus [$\rho(0)$] and the other expressions derived using the uniform, Gaussian, and Fermi charge distribution models.
These results are presented in Table~\ref{tab9}. From this table, we find that the FS factors vary little with different nuclear charge distributions, and the differences can be neglected compared to other sources of uncertainty.
Since the Fermi charge distribution is a more realistic model compared to other models under consideration, we use FS factors determined using this distribution in the FF approach as our final results.   

\section{Testing calculations with muonic and electronic ISs}

In this section, we consider our calculated IS factors in light of the available experimental data.
As the NMS factor values obtained with the FF approach agree with those calculated with the scaling law for all states, and as the relativistic and electron correlation effects are small and well converged, we deem these factors reliable.
The FS factors agree to within a few percent between approaches, with prior calculations, and with their semi-empirical values based on the hyperfine factors. They may thus be taken from any method, and here we take the values calculated with the FF approach.
We consider the two best strategies for obtaining the total MS. The first is to add the NMS factor from the scaling law to the calculated SMS factor via the AR approach, as was done in~\cite{Koszorus}. This has the advantage that possible higher-order contributions, which may be present when using the FF approach, are absent. The second is to add the calculated NMS and SMS factors from the FF approach. For transitions, both strategies result in similar MS factors although their correlation trends are different, making the results independent.

\begin{table}[htbp]
\caption{
Comparison of the calculated and semi-empirically (SE) estimated MS factors. The first column lists the experimentally considered transitions in shorthand notation. The second column is the weighted average center-of-gravity $^{39,41}$K measured ISs for each transition from Refs.~\cite{1981-Bend, 2006-Falke,2009-Behr, 2011-Kohl, 2011-46}. $K_\text{SE}^\text{tot}$ is the semi-empirical MS factor calculated from the measured ISs, $F$ from Table~\ref{tab:FF} and $\langle r_N^2 \rangle^{39,41}=0.119(27)~$fm$^2$ from the muonic atom (see main text). Comparison between the last two columns shows that $K_\text{SE}^\text{tot}$ agrees well with our estimation for all the above transitions when IS factors used from the FF approach and less so when they are calculated using the AR approach. 
}
\begin{ruledtabular}
\begin{tabular}{l c | cc c}
  &  IS  & $K^\text{tot}_\text{SE}$ & $K^\text{tot}_\text{FF}$ & $K^\text{SMS}_\text{AR}$+$K^\text{NMS}_\text{SE}$\\
  &  MHz &   GHz~u & GHz~u  & GHz~u
\\
\hline \\
$4S-4P_{1/2}$  & 235.5(1) & 198.5(2.4) & 194.3(3.3) & 193.7(4.7) \\
$4S-4P_{3/2}$  & 236.2(2) & 199.0(2.4) & 197.2(3.4) & 198.9(4.6) \\
$4S-5P_{1/2}$  & 455.2(9) & 373.6(2.5) & 374.5(5.5) & 381.4(5.9) \\
$4S-6S$        & 500.8(6) & 408.9(2.1) & 411.2(6.0) & 420.6(5.6) \\

\end{tabular}
\end{ruledtabular}
\label{ISdata}
\end{table}
Benchmarking against experimental values could allow us to distinguish which strategy is more robust in determining MS factors. To do so, one could in principle combine the measured ISs, masses, and charge-radii of at least two isotope pairs and deduce experimental IS parameters directly from Eq.~\ref{IS}. This method is referred to as a Calibrated King plot. However, independent absolute charge radii measurements have only been achieved with the two stable K isotopes~\cite{1981-WSH}, while the radii of at least three isotopes are needed for a calibrated King plot.

In light of this, we adopt a different method, whose results are given in Table~\ref{ISdata}. 
It relies on $\langle r_N^2 \rangle^{39,41}$ measured with muonic atoms~\cite{1981-WSH} combined with our calculated FS factor. The uncertainty reported in Ref.~\cite{1981-WSH} stems from the measured muonic atom energies and a possible model dependence of the differential charge distribution. It does not allow uncertainty in nuclear polarization~\cite{1981-WSH}. 
In the next section, we update the nuclear polarization calculation and show that although the absolute radii change significantly, $\delta r^{39,41}_N$ is nearly unaffected.
Our recommended value employing the updated nuclear polarization, allowing for uncertainty in it, is $\delta r^{39,41}_N=17.3(4.0)~$am.
The corresponding mean square difference is $\langle r_N^2 \rangle^{39,41}=0.119(27)~$fm$^2$, which is close to the value given in~\cite{Martensson2}.

For each transition, the semi-empirical total MS factor, $K_\text{SE}^\text{tot}$ given in Table~\ref{ISdata}, is extracted using $\langle r_N^2 \rangle^{39,41}$ and the FS factors from Table~\ref{tab:FF}.
As seen in Table~\ref{ISdata}, the resulting $K_\text{SE}^\text{tot}$ agree for all transitions with those calculated with the FF approach to within than their combined errors. Compared with the MS factors obtained by summing $K^\text{SMS}_\text{AR}$ and $K^\text{NMS}_\text{SE}$, we see agreement for two transitions but a deviation of two combined errors for the other two. 
We surmise that for these transitions in K, the strategy of obtaining the MS factor from the FF calculations is preferable.

\section{Updated radius of $ ^{38m}$K}

The center of gravity IS between $^{38m}$K and $^{39}$K was measured for the $4S-4P_{3/2}$ transition using a magneto-optical trap~\cite{19973738KD2}.
Later, ISs for the $4S-4P_{1/2}$ transition for pairs $^{38m,38}$K and $^{38m,39}$K were measured with collinear laser spectroscopy~\cite{201438K}. 
We consider these results on their own, and also as combined with other IS measurements for the $^{38,39}$K pair~\cite{1982-38-47K, 2019-K3847}.
This results in four, almost uncorrelated, IS values given in Table~\ref{tab:38}.

Our final recommended values for the total MS factors, $K_\text{WA}^\text{tot}$, are the weighted averages of $K_\text{SE}^\text{tot}$ and $K_\text{FF}^\text{tot}$. From them and from the FS factors of Table~\ref{tab:FF}, we extract the mean squared radius difference $\langle r_N^2\rangle^{38m,39}_\text{WA}$. The results are given in Table~\ref{tab:38}, where it is seen that the $\langle r_N^2\rangle^{38m,39}_\text{WA}$ extracted from different experiments agree within their nearly uncorrelated experimental uncertainty, indicating consistency. We can thus report their average, weighted by the experimental uncertainty, as a final value. The systematic uncertainty resulting from $K_\text{WA}^\text{tot}$ is added separately.
Our final result is $\langle r_N^2\rangle^{38m,39}_\text{WA}=-0.019(14)~$fm$^2$, where the uncertainty includes both correlated and uncorrelated sources. It agrees with and is twice as accurate as the value quoted in the previous work~\cite{201438K}. The improvement comes primarily from our updated calculation of the IS factors.

To determine the absolute radius, one should use a reference isotope. The best choice is $^{39}$K, which was subject to both muonic atom x-ray spectroscopy~\cite{1981-WSH},  and high-statistics, broad momentum transfer electron scattering~\cite{1973-Scat}, providing a charge distribution that is nearly model independent.
The radius obtained using this distribution and with the nuclear polarization correction $\Delta E_\text{NP}=119(36)~$eV tabulated in~\cite{2004-FH}, is $r_N(^{39}\text{K})=3.4353(29)~$fm~\cite{2024-Mirr}.
In this work, we revisit this correction. 
We find $\Delta E_\text{NP}=156(47)~$eV, which is $37~$eV higher than the value reported in~\cite{2004-FH}.

The main effect missing in all previous estimates is that due to virtual excitations in the hadronic range, which can be coined ``nucleon polarization''. The latter contribution has been extensively studied for the lightest muonic systems ($\mu$H,\,$\mu$D,\,${\mu}^{3,4}$He$^+$, see~\cite{2024-RMP} and references therein).
Details of the updated calculation of the nuclear polarization, including the aforementioned effect of the nucleon polarization, will be published elsewhere.

This adjustment exceeds both the experimental uncertainty of $32$~eV and the $36$~eV uncertainty attributed to nuclear polarization by the authors of Ref.~\cite{2004-FH}.
Because the finite-size effect reduces the transition energy, the charge radius must be larger to compensate for this change. Adopting $C_z=-0.050$~am/eV from~\cite{2004-FH} we obtain a correction of $1.85~$am to the Barret-equivalent radius and $1.44~$am to the charge radius. 
Thus, we can recommend $r_N(^{39}\text{K})=3.4367(31)~$fm, with uncertainty dominated by the nuclear model and polarization.

Combining the differential and reference radii returns $r_N(^{38m}\text{K})=3.4396(37)~$fm.
This may be compared to the previous value used in determining $V_{ud}$, $r_N(^{38m}\text{K})=3.437(4)~$fm~\cite{201438K}. 
Our somewhat larger radius, by approximately one standard error, is derived from the cumulative shifts caused by the IS factors and the nuclear polarization corrections derived in this study, both of which add constructively.
The similar uncertainty results from the decrease in the uncertainty in $\langle r_N^2\rangle^{38m,39}$ from the more accurate calculation of the MS factor, balanced by the more conservative estimate of the uncertainty in $r_N(^{39}\text{K})$.
Our recommended radius is a crucial ingredient in the extraction of the $V_{ud}$ matrix element from super-allowed beta decay $^{38m}\text{K}\rightarrow~^{38}\text{Ar}$~\cite{Seng:2023cgl}.


\begin{table}[htbp]
\caption{
Comparisons of differential radius extraction for the pair $^{38m,39}$K. The first column lists the experimentally considered transitions in shorthand notation. The second column is the center-of-gravity $^{39,40}$K ISs from combinations of the references indicated in the third column. The fourth column lists the MS factors that are the weighed average of columns 3 and 4 of Table~\ref{ISdata}. The last column is the extracted mean-square radius differences using these MS factors, and the FS factors of Table~\ref{tab:FF}. The nearly uncorrelated statistical uncertainty from the IS measurements is given in parenthesis and the highly correlated uncertainty from the atomic factor calculations are given in square brackets. The weighted average is reported below and compared with the prior works.
}
\begin{ruledtabular}
\begin{tabular}{l cl cr}
  &  IS  & Ref. & $K_\text{WA}^\text{tot}$ & $\langle r_N^2 \rangle^{38m,39}_\text{WA}$ \\
 & MHz &&GHz~u& fm$^2$ \\
\hline \\
$4S-4P_{1/2}$  & $134.5(1.1)$ & \cite{201438K}                   & $197.0[19]$& $-0.018(10)[12]$  \\
\hspace{10 pt} "  & $136.4(2.1)$ & \cite{201438K}+\cite{2019-K3847} & " & $-0.035(19)[12]$  \\
\hspace{10 pt} " & $138.0(5.3)$ & \cite{201438K}+\cite{1982-38-47K} & " & $-0.050(48)[12]$  \\
$4S-4P_{3/2}$  & $132.0(3.0)$ & \cite{19973738KD2}               & $198.4[20]$& $~0.013(27)[12]$  \\\\

\multicolumn{2}{l}{Weighted Average} && \multicolumn{2}{r}{$-0.019(08)[12]$}\\
\multicolumn{2}{l}{Previous Work} & \cite{201438K} && $-0.011(10)[23]$ \\
\end{tabular}
\end{ruledtabular}
\label{tab:38}
\end{table}

\section{Testing for isospin-symmetry breaking}

In the ongoing effort to obtain $V_{ud}$ from superallowed beta decays of nuclear isotriplets, charge radii are essential to evaluate theoretical calculations. They provide a crucial benchmark for the isospin symmetry breaking correction $\delta_\text{C}$, which modifies $|M_F|^2$, the square of the Fermi matrix element, from its isospin limit value $|M_F^0|^2=2$.
Within a nuclear model (for example, Woods-Saxon~\cite{Towner:2002rg,Towner:2007np,Hardy:2008gy,Xayavong:2017kim}), the uncertainty in $\delta_C$, resulting mainly from the uncertainty of the radius parameter in the model, is typically small; however, a considerable variation of the results of $\delta_\text{C}$ between different models is observed~\cite{Ormand:1989hm,Ormand:1995df,Satula:2011br,Satula:2016hbs,Liang:2009pf,Auerbach:2008ut}. It is thus necessary to differentiate between these models, and 
Ref.~\cite{Seng:2022epj,Seng:2023cvt} demonstrated that it can be done by examining the models' capability to calculate a correlated isospin-symmetry-breaking quantity, which is derived from a combination of charge radii of the nuclear isotriplet:
\begin{equation}\label{eq:MB}
    \Delta M_B^{(1)}\approx\frac{1}{2}\left( Z_{+1} r^2_{N,+1}+Z_{-1} r^2_{N,-1} \right)-Z_0 r^2_{N,0}~,
\end{equation}
where the subscript $\pm 1,0$ are the eigenvalues of the isospin operator $T_3$ of the nucleus. It provides a rare opportunity to directly compare theoretical studies of isospin-breaking effects to experiments, since the charge radii are measurable quantities.
We note here that the original Eq.~(10) in Ref.~\cite{Seng:2022epj} deals with the point-proton radii, but that the proton, neutron and Darwin-Foldi contributions, which translate the point-proton to the charge radii, cancel in the combination, so that the approximation of Eq.~(\ref{eq:MB}) is due only to neglecting the residual spin-orbit effects.

Plugging in the values for the $^{38}$Ca, $^{38m}$K and $^{38}$Ar radii to Eq.~(\ref{eq:MB}) results in a value consistent with zero, but with too large of an uncertainty to test isospin symmetry breaking. This may be traced back to the missing electron scattering experiments in $^{38}$Ar~\cite{2024-Mirr}. 
To perform a more stringent test, we assume that the mirror fit holds in this region of the nuclear chart, so that in general $r_{N,-1}=r_{N,+1}+I\times1.382(34)~$fm~\cite{2024-Mirr}, and specifically $r_N(^{38}\text{Ca})=r_N(^{38}\text{Ar})+0.0727(18)~$fm.
This enables to compute $\Delta M_B^{(1)}(38)$ using the two better known charge radii of the triplet, namely $r_N(^{38}\text{Ca})$ and $r_N(^{38m}\text{K})$.

Before doing so, we revisit the radius of $^{38}$Ca.
First, we take $\langle r_N^2 \rangle^{38,40}_\text{Ca}=0.0797(64)~$fm$^2$ as measured by collinear laser spectroscopy~\cite{Miller}. Its uncertainty is completely dominated by the calibration of the beam energy~\cite{Miller}.
Next, we update the reference radius of $^{40}$Ca, from $r(^{40}\text{Ca})=3.4807(28)~$fm~\cite{2024-Mirr}, derived with $\Delta E_\text{NP}=142~$eV from Ref.~\cite{2004-FH}, to 
$r(^{40}\text{Ca})=3.4818(30)~$fm~\cite{2024-Mirr}, re-derived here with $\Delta E_\text{NP}=177~$eV which includes the nucleon polarization contributions.
The updated value now becomes $r(^{38}\text{Ca})=3.4703(31)~$fm, resulting in $r(^{38}\text{Ar})=3.3973(36)~$fm which is compatible but twice as accurate as the value $r(^{38}\text{Ar})=3.402(6)~$fm~\cite{2024-Mirr}.

Applying Eq.~(\ref{eq:MB}) results in $\Delta M_B^{(1)}(38)=-0.48(63)~$fm$^2$, where correlations have been taken into account in the uncertainty estimation \footnote{Note that at the current precision level $\Delta M_B^{(1)}(A=38)$ is only slightly affected by the updated $\Delta E_\text{NP}$ since the shift has the same sign and nearly identical size for all members of the isotriplet.}.
$\Delta M_B^{(1)}(A=38)$ is consistent with zero within its uncertainty, as implied indirectly by various nuclear models~\cite{Seng:2022epj}. 
However, the same models also suggest a small $\Delta M_B^{(1)}(A=26)$, deviating from its semiempirical value by staggering $4.5~\sigma$~\cite{2024-Mirr}. 
This motivates us to revisit the extraction of the $^{26m}$ Al charge radius, which we reserve for future work.

\section{Experimental prospects}

This study highlights the significance of integrating muonic and electronic atom spectroscopy across multiple transitions to develop a database of semi-empirical atomic factors, which can then be used to test many-body atomic theory calculations.
Consequently, a better understanding of the uncertainties in atomic theory would significantly impact the determination of nuclear radii across the chart~\cite{review}. This effect is especially notable in mono-isotopic elements, where a (full or partial) calibrated King Plot procedure is not applicable.

In K, enhancing the precision of $K_\text{SE}^\text{tot}$ for the transitions that have already been measured would require a more accurate determination of $\langle r_N^2 \rangle^{39,41}$ from muonic atoms. Currently, it is limited by our knowledge of the charge distribution of $^{41}$K, which could either be measured with electron scattering as in $^{39}$K~\cite{1973-Scat}, or calculated using ab initio nuclear many-body methods~\cite{1975-Friar,2021-Density}.
An improved knowledge of the charge distribution would open the door for a more precise measurement of muonic atoms and a microscopic calculation of nuclear polarization.

Another promising avenue to improve semiempirical mass shift extraction is to measure $^{39,40}$K with muonic atoms. 
This endeavor would be particularly challenging as $^{40}$K is not naturally abundant, so it must be measured in microscopic quantities, a feat that has only recently been approached at the PSI accelerator~\cite{2023-MuX1,2023-MuX2,2024-MuX}.
Despite the experimental complexity, if such a measurement could be performed, then the similarity in radii between $^{39}$K and $^{40}$K would considerably reduce the uncertainty of the theory in their difference, making them highly suitable for extraction of improved IS factors.
Ideally, this measurement would be coupled with electron scattering from a microscopic sample of $^{40}$K, which could be performed at the SCRIT facility~\cite{2017-SCRIT}.

We now consider some prospects for improved optical IS measurements.
Long-lived isotopes of K can be subjected to laser cooling and trapping in an atomic state, allowing for measurements with precision that is a fraction of the natural linewidth.
Our calculations indicate that the $5S$ state is reasonably sensitive to the nuclear charge radius, while higher-order electron correlations have much less impact on it compared to the ground state.
It would thus be interesting to extend measurements with long-lived isotopes to transitions which include the $5S$ state.
Suitable one-photon transitions are $4P_{1/2}-5S$ ($4P_{3/2}-5S$) around $1250~$nm, with a linewidth of $6$ $(4)$~MHz. Another option is to perform two-photon spectroscopy of the $4S-5S$ transition at roughly $2\times950$~nm, as was done for the $4S-6S$ interval~\cite{2011-46, 2017-Comb}.
Once at least two isotope shifts are well measured for an interval for which the calculated IS factors are well converged, then they can be projected to other transitions using (uncalibrated) King Plots (see e.g.~\cite{2022-CdHan,2023-Zn}).

For rare isotopes, and particularly $^{38{\rm m}}$K, statistical accuracy plays a major role, so we consider single-photon transitions from the ground state.
Here, the $4s~^2S_{1/2}\rightarrow5p~^2P_{1/2}$ line is a promising candidate due to  
its natural linewidth of $\Gamma=1.1$~MHz, narrower than $\Gamma=6$~MHz of the $4P_{1/2}$ and $4P_{3/2}$ states used in previous measurements~\cite{19973738KD2, 201438K}.
To fully exploit this narrow linewidth, a Doppler-free measurement scheme is needed, such as performing measurements in a magneto-optical trap (MOT).
The minimal lifetime needed is roughly $1$~s (see Table~I in~\cite{2020-MOTS}) spanning $37-49~$K.
Techniques developed to spin-polarize beta decaying isotopes with the MOT off, including coherent population trapping~\cite{Gu2003}, could be used to minimize Zeeman shifts by quickly turning off the magnetic field of the MOT~\cite{Fenker2014}.

\section{Summary}

We have performed a detailed analysis of the roles of electron correlation effects in the determination of isotope shift factors in a few low-lying states of the potassium atom.
Relativistic coupled-cluster theory in the singles, doubles, and triples approximation was employed for this purpose.
Our results were derived using finite-field methods, expectation value calculations, and analytical response approaches.
The differential values for the field-shift and specific mass-shift factors showed reasonable agreement between the finite-field and analytical response methods, but not for the normal mass-shift factors.
This deviation is attributed to orbital relaxation effects, which are considered only perturbatively in the analytical response approach and may significantly influence the normal mass-shift factors.
Evaluating and comparing these methodologies, as well as contrasting them with semi-empirical data, revealed that the finite-field method provides the most accurate results for potassium.
The expectation value method proved less reliable for determining the normal and specific mass shift factors. However, it allowed us to verify a useful empirical correlation between magnetic dipole hyperfine structure- and field-shift factors, which may be advantageous to apply to heavier systems.

Our final recommended values for the mass shift factors were determined as a weighted average of those we calculated using the finite-field method and those that we extracted semi-empirically from muonic atom measurements. These are of similar accuracy to the ones used in~\cite{Koszorus} but have a much more robust and transparent uncertainty estimation. This highlights the difficulty in assessing realistic uncertainties in many-body calculations of isotope shift factors and motivates a new generation of muonic atom x-ray spectroscopy, analyzed with modern tools, to better test the atomic theory.

The weighted average mean-square radius difference $\langle r_N^2\rangle^{38m,39}_\text{WA}=-0.019(14)~$fm$^2$ is extracted consistently from the available experimental results and our recommended isotope shift factors. It is twice as accurate as the previous result given in the literature~\cite{201438K} and has a substantial experimental uncertainty, opening up the opportunity for improved optical measurements of higher precision. We have briefly mentioned several suggestions, informed by the uncertainty in the theory calculations, to improve this precision.

Combining $\langle r_N^2\rangle^{38m,39}_\text{WA}$ with the reference radius of $^{39}$K, we obtained an updated $r_N(^{38m}\text{K})=3.4396(37)~$fm, which differs by one standard error from the previously recommended value given in Ref.~\cite{2024-Mirr}.
The reason for the deviation is that we account for a previously overlooked nucleon polarization correction to the macroscopic nuclear polarization calculation in medium mass muonic atoms. Details on the new calculation will be given in a separate publication.

Finally, we combine $r_N(^{38m}\text{K})$ with a similarly revised $r_N(^{38}\text{Ca})$ and the results given in Ref~\cite{2024-Mirr} to obtain the isospin symmetry breaking matrix element $\Delta M_B^{(1)}(38)=-0.48(63)~$fm$^2$. It is found to be consistent with zero within its uncertainty, constituting the most stringent test of isospin symmetry breaking using charge radii~\cite{2024-Mirr}.

Various nuclear models also imply $-0.42$~fm$^2<\Delta M_B^{(1)}(38)<-0.04$~fm$^2$~\cite{Seng:2022epj} in agreement with our result. However, the same models suggest $-0.12$~fm$^2<\Delta M_B^{(1)}(26)<-0.03$~fm$^2$~\cite{Seng:2022epj} in stark contrast to the results given in Ref.~\cite{2024-Mirr}.
This motivates us to revisit all aspects of the extraction of the radii of the $A=26$ isotriplet which we reserve for future work.

\begin{acknowledgments}
B.K.S. acknowledges ANRF for grant no. CRG/2023/002558. VK, AC and BKS were supported by the Department of Space, Government of India, to carry out this work. All atomic calculations reported in the present work were computed using the ParamVikram-1000 HPC cluster of the Physical Research Laboratory (PRL), Ahmedabad, Gujarat, India. M.~G. acknowledges support by EU Horizon 2020 research and innovation programme, STRONG-2020 project under grant agreement No 824093, and by the Deutsche Forschungsgemeinschaft (DFG) under grant agreement GO 2604/3-1. The work of C.-Y.S. is supported in part by the U.S. Department of Energy (DOE), Office of Science, Office of Nuclear Physics, under the FRIB Theory Alliance award DE-SC0013617, and DOE grant DE-FG02-97ER41014, and he also acknowledges support from the DOE Topical Collaboration ``Nuclear Theory for New Physics'', award No. DE-SC0023663.
B.O. is grateful for the support of the Council for Higher Education
Program for Hiring Outstanding Faculty Members in Quantum Science and Technology.
\end{acknowledgments}

\bibliography{references}

\end{document}